\documentclass{egpubl}
 
\JournalSubmission    
\usepackage[T1]{fontenc}
\usepackage{dfadobe}  

\usepackage{cite}  
\BibtexOrBiblatex
\electronicVersion
\PrintedOrElectronic
\ifpdf \usepackage[pdftex]{graphicx} \pdfcompresslevel=9
\else \usepackage[dvips]{graphicx} \fi

\usepackage{multirow}
\usepackage{multicol}
\usepackage{egweblnk} 

\usepackage{amsmath}
\usepackage{amssymb}
\usepackage{amsfonts}
\usepackage{mathtools}

\usepackage{caption}
\usepackage{subcaption}
\usepackage[ruled,vlined]{algorithm2e}
\usepackage[dvipsnames]{xcolor}
\usepackage{float}
\usepackage{xspace}

\usepackage{soul}

\newcommand{\myparagraph}[1]{\vspace{1mm} \noindent \textbf{#1}}
\newcommand{\parext}{\textsc{tachyon}\xspace}
\newcommand{\gmsc}{g\textsc{msc}\xspace}

\newcommand{\highlightadd}[1]{{#1}\xspace }

\title[\parext : Efficient Shared Memory Parallel Computation of Extremum Graphs]%
      {\parext : Efficient Shared Memory Parallel Computation of Extremum Graphs}

\author[A. Ande, V. Subhash \& V. Natarajan]
{\parbox{\textwidth}{\centering 
    Abhijath Ande\orcid{0000-0003-0575-9350}, 
    Varshini Subhash\orcid{0000-0003-1889-6821}, and
    Vijay Natarajan\orcid{0000-0002-7956-1470}
    }
    \\
{\parbox{\textwidth}{\centering Department of Computer Science and Automation, \\ Indian Institute of Science, Bangalore, India}
}
}

\begin{document}


\maketitle
\begin{abstract}
The extremum graph is a succinct representation of the Morse decomposition of a scalar field. It has increasingly become a useful data structure that supports topological feature directed visualization of 2D / 3D scalar fields, and enables dimensionality reduction together with exploratory analysis of high dimensional scalar fields. Current methods that employ the extremum graph compute it either using a simple sequential algorithm for computing the Morse decomposition or by computing the more detailed Morse-Smale complex.  Both approaches are typically limited to two and three dimensional scalar fields. We describe a GPU-CPU hybrid parallel algorithm for computing the extremum graph of scalar fields in all dimensions. The proposed shared memory algorithm utilizes both fine grained parallelism and task parallelism to achieve efficiency. An open source software library, \parext, that implements the algorithm exhibits superior performance and good scaling behavior.

\ccsdesc[300]{Human-centered computing~Visualization techniques}
\ccsdesc[500]{Computing methodologies~Parallel algorithms}

\printccsdesc   
\end{abstract}  
\section{Introduction} 
\label{sec:introduction}

Topological analysis has become a key tool for the analysis and visualization of data from various branches of scientific research~\cite{carr2020topological,hotz2021topological}. In the context of data represented as scalar fields, topological descriptors such as contour trees, Reeb graphs, merge trees, Morse-Smale complexes, and extremum graphs have been widely studied~\cite{heine2016survey}. Each abstraction provides a different perspective to analyze and interact with the scalar field. The extremum graph is a useful abstraction in situations where the extrema and gradient flow behavior that determines their connectivity are crucial for understanding the scientific phenomenon~\cite{correa2011topological, thomas2013detecting}. It is a subset of the Morse-Smale complex that is considerably simpler yet intuitive and detailed enough for various applications. It is a bipartite graph whose nodes are either extrema or saddles and arcs determine connectivity via a gradient path between a saddle-extremum pair.  The simple structure of the graph makes it amenable for various applications including segmentation in 2D and 3D scalar fields, feature tracking in time-varying data, and clustering in higher dimensional data. While there exist efficient algorithms to compute the more general Morse-Smale complex, there is a need for a lightweight, fast, and scalable algorithm that can compute the extremum graph in all dimensions. 

\subsection{Related work}

The Morse complex and Morse-Smale complex were introduced as topological structures to represent the gradient flow behavior of Morse functions~\cite{bremer2004topological,edelsbrunner2003morse}. While the extremum graph may be considered as a 1-skeleton of the Morse complex, it was first described in the currently known form by Correa et al.~\cite{correa2011topological} together with a new 2D visual representation called the topological spine. The extremum graph preserves the topological and geometric structure of a scalar field, while the topological spine provides a succinct and easily digestible 2D representation to the user. It essentially preserves the relative location of extrema and knowledge of their neighborhoods with respect to gradient paths connecting them. Useful applications of extremum graphs include the robust detection of symmetry in noisy scalar fields~\cite{thomas2013detecting}, extrema proximity awareness and a feature-aware comparison of similar scalar fields using distance measures~\cite{narayanan2015distance}. 

The abundance of large-scale scientific data with increasing feature complexity and precision has directed the attention of the scientific visualization community towards developing parallel algorithms that effectively leverage modern compute power and massively parallel architectures. Specifically, the parallel computation of 3D Morse-Smale complexes has been studied extensively to this effect. Gyulassy et al.~\cite{gyulassy2008} developed a memory efficient algorithm for 3D Morse-Smale complex computation, where they partitioned large data into chunks called parcels that fit in memory. By computing the Morse-Smale complexes for each parcel and performing a cancellation based merging of parcels, they were able to obtain Morse-Smale complexes for large-scale data which originally did not fit in memory. This approach was adapted to a distributed memory setting by Peterka et al.~\cite{peterka2011scalable} and Gyulassy et al.~\cite{gyulassy2012parallel} where they leveraged massively parallel clusters to handle the parcels in parallel. Robins et al.~\cite{robins2011theory} and Shivashankar and Natarajan~\cite{shivashankar2011parallel} introduced novel locally independent definitions for gradient pairs which facilitated an embarrassingly parallel gradient pair assignment. Novel traversal algorithms for the extraction of ascending and descending manifolds of the extrema and saddles further enhanced the runtime performance of the Morse-Smale complex computation~\cite{gunther2012efficient, shivashankar2012parallel3d, delgado2014skeletonization}. 

Additionally, there have been advances in obtaining a higher degree of geometric accuracy during parallel Morse-Smale complex computation by Gyulassy et al.~\cite{gyulassy2018shared, gyulassy2014conforming, gyulassy2012computing} and Bhatia et~al.~\cite{bhatia2018topoms}. Several approaches dabbled in CPU based shared memory parallelization techniques with the exception of Shivashankar and Natarajan~\cite{shivashankar2012parallel3d}, who introduced a hybrid approach. Their algorithm and the associated software library, pyms3d~\cite{mscsoftware2017}, leveraged GPU parallelism for gradient assignment and extrema traversal and CPU parallelism for saddle-saddle traversal. Recently, Subhash et al.~\cite{subhash2020GPU,subhash2022gpu} introduced \gmsc, an end-to-end GPU parallel algorithm for 3D Morse-Smale complex computation, which resulted in substantial speedups over the CPU and hybrid approaches attempted thus far. \highlightadd{\gmsc transforms the saddle-saddle computation into a sequence of matrix operations that are amenable to fast parallel computation on the GPU, leading to significant improvement in runtime. We note that the gradient assignment and saddle-extrema traversal steps in \gmsc are identical to those employed by pyms3d. Both pyms3d and \gmsc are restricted to 3D scalar fields.}

A discrete Morse theory based approach is employed by several of the above-mentioned parallel algorithms and has yielded combinatorial and numerically robust algorithms~\cite{gyulassy2006simplification,gunther2012efficient,delgado2014skeletonization,shivashankar2011parallel,shivashankar2012parallel3d}. Fugacci et al.~\cite{fugacci2019computing} also demonstrated an extension to higher dimensions by computing the Morse complex from simplicial complexes. This discrete Morse complex can be leveraged to compute homology~\cite{robins2011theory,harker2014discrete}, perform shape analysis and study scalar fields~\cite{defloriani2015morse}. 

As mentioned earlier, the extremum graph is a subgraph of the combinatorial structure or the 1-skeleton of the Morse-Smale complex. The succinct yet highly informative abstraction and its demonstrated role in many applications motivates the need for its efficient computation. To the best of our knowledge, no previous work specifically explores the efficient and scalable parallel computation of extremum graphs in all dimensions. We attempt to address this gap here and describe an open source implementation.

\subsection{Contributions}
In this paper, we describe a fast shared memory parallel algorithm for computing the extremum graph of an $n$-dimensional scalar field defined on a grid. The hybrid GPU-CPU parallel computation leverages the GPU for fine-grained parallel computation of nodes of the extremum graph and the CPU for effective task parallel computation of arcs of the graph. Key contributions of the paper include:
\begin{itemize}
\item A hybrid GPU-CPU algorithm that efficiently uses both computational resources while enabling the algorithm to be scalable to data that does not fit in the GPU memory. 
\item An efficient implicit representation of the neighborhood of a grid vertex that supports fast parallel critical point classification of all vertices. 
\item \parext, a lightweight library that provides a fast implementation of the hybrid algorithm~(\href{https://bitbucket.org/vgl_iisc/tachyon}{\url{bitbucket.org/vgl_iisc/tachyon}}).
\item Multiple strategies for effective simplification of the extremum graph that makes it amenable for feature identification and applications.
\item Experimental studies on multiple datasets to demonstrate superior performance and scalability. 
\end{itemize}

\section{Background}
\label{sec:background}
In this section, we define and briefly introduce the necessary terms required to define extremum graphs~\cite{edelsbrunner2003morse,correa2011topological,narayanan2015distance}. Given a smooth function $f: \mathbb{M} \rightarrow \mathbb{R}$ over a  manifold $\mathbb{M}$ of dimension $n$, we say that a point $x \in \mathbb{M}$ is \emph{critical} iff $\nabla f(x) = 0$, $x$ is called \emph{regular} otherwise. Also, $f$ is a \emph{Morse function} if all critical points have pairwise distinct function values and none of them are degenerate {i.e.}, the Hessian evaluated at the point is non-singular. For a critical point $x$ of $f$, its Morse index is defined as the number of negative eigenvalues of its Hessian matrix evaluated at $x$. Critical points of index $0$ are named \emph{minima}, index $n$ are \emph{maxima} and the others with index $k$, $1<k<n-1$, are called $k$-\emph{saddles}.

An \emph{integral line} is a maximal curve in $\mathbb{M}$ whose tangent at every point is equal to the gradient of $f$ at that point. Naturally, $f$ monotonically increases along the integral line and its two end points (limit points) are critical points of $f$. The Morse function $f$ determines a decomposition of $\mathbb{M}$  based on the integral lines. The union of integral lines that terminate at a critical point is called its \emph{descending manifold}. The descending manifold of a maximum is an $n$-dimensional manifold. The collection of descending manifolds of critical points of $f$ partition $\mathbb{M}$ and is called the \emph{Morse decomposition}. The \emph{ascending manifold} of a minimum is similarly defined as the union of integral lines that originate at a minimum. Again, the collection of ascending manifolds of critical points of $f$ partition $\mathbb{M}$. The \emph{extremum graph} is a  representation of the Morse decomposition. It is called a maximum / minimum graph if it represents the decomposition into descending / ascending manifolds, respectively. Without loss of generality, we restrict the discussion in this paper to maximum graphs, and refer to them as extremum graphs. 

\begin{figure}
    \centering
    \includegraphics[scale=0.12]{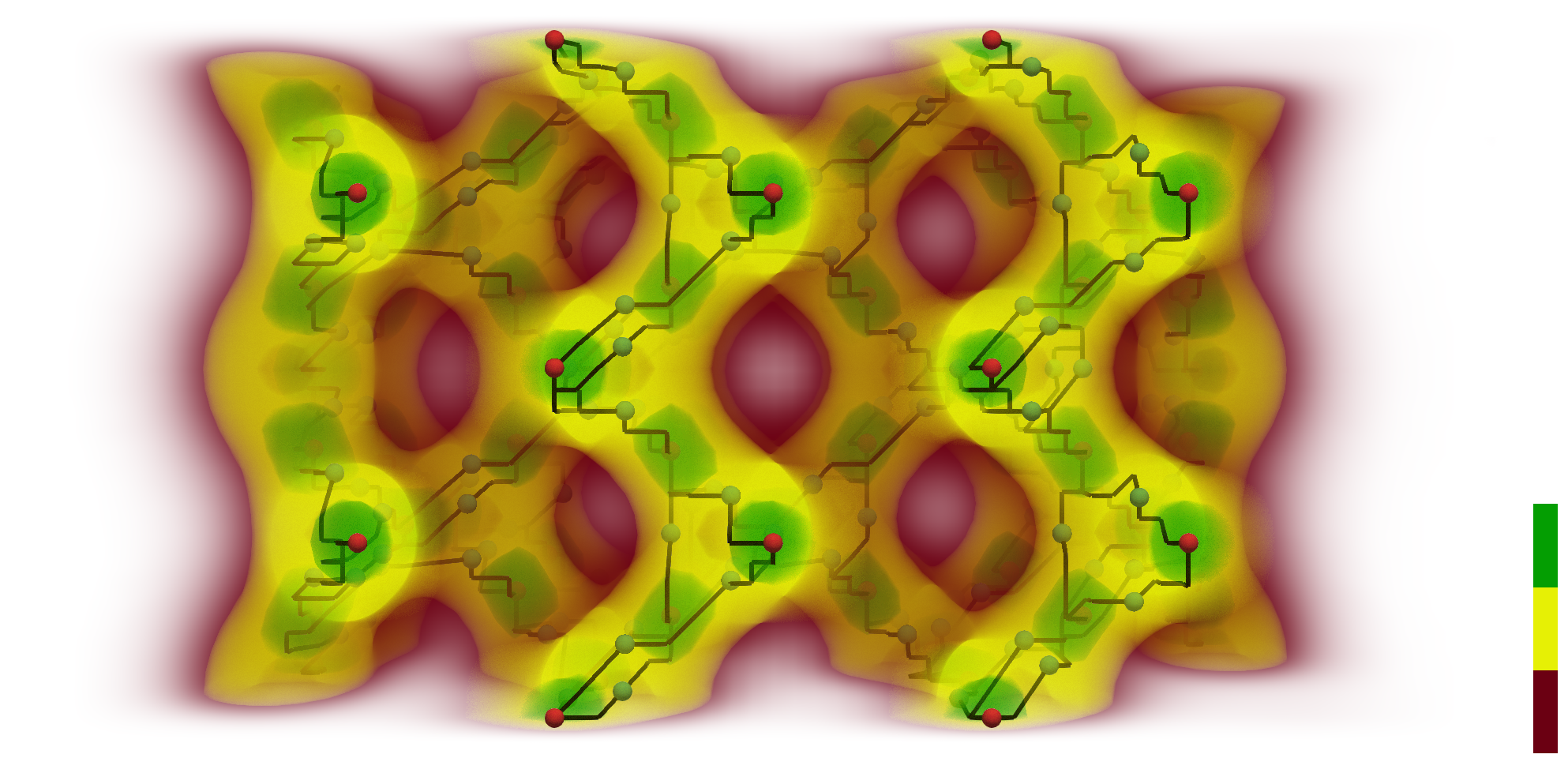}
    \caption{Extremum graph for the silicium grid. Maxima (red) and 2-saddles (green) are connected by arcs. The geometry of an arc is determined by the integral line (black) that connects the maximum and the 2-saddle. The scalar field is shown as a volume rendering using a simple banded color map.} 
    \label{fig:extremumgraph-illustration}
\end{figure}

An $(n-1)$-saddle $s$ of $f$ lies on the boundary of the descending manifold of a maximum $m$ {i.e.}, an integral line originating at $s$ terminates at $m$. The extremum graph captures this  relationship between maxima and $(n-1)$-saddles of $f$, and thereby captures the combinatorial structure of the Morse decomposition. The node set of the extremum graph consists of the maxima and $(n-1)$-saddles of $f$.  An arc $(s,m)$ belongs to the extremum graph if $s$ lies on the boundary of the descending manifold of $m$. Figure~\ref{fig:extremumgraph-illustration} shows the extremum graph for a 3D scalar field, the output of a simulation of a silicium grid.

\section{Computation of extremum graphs}
\label{sec:algorithm}

There are two popular approaches to compute extremum graphs, namely flood fill and gradient path tracing. While both methods produce equivalent results, each method has its own associated advantages and disadvantages. We begin with a brief discussion about the two approaches and justify why we propose gradient path tracing towards the development of an efficient and scalable parallel algorithm. 

\subsection{Flood fill \textit{vs.} gradient path tracing}
The flood fill approach computes the extremum graph in one sweep over the list of grid vertices by incrementally growing the descending $n$-dimensional manifolds of all maxima. The vertices are processed in decreasing order of scalar value. Processing a vertex $v$ includes expanding the descending manifolds that have reached $v$ in the steepest descent direction, namely the vertex with the smallest value in the lower link. This approach has been employed to compute the Morse decomposition~\cite{edelsbrunner2003morse}. The method implicitly recognizes a maximum as a vertex where no descending manifold have reached and an $(n-1)$-saddle as a vertex where multiple descending manifolds merge. The adjacency between maxima and saddles is recorded when the saddle vertex is processed.  While this method computes the connectivity of the extremum graph, it does not explicitly capture the gradient paths between the maxima and saddles. A subsequent step needs to be invoked to compute these gradient paths as necessary, a limitation of this approach. Further, this method has a high memory footprint because it requires the collection of all descending manifolds to be stored. 

The gradient path tracing approach computes the extremum graph in two steps. The first step locates and classifies the maxima and $(n-1)$-saddles. It analyzes the link of a vertex to classify the vertex. Importantly, the classification of a vertex is independent of the classification of other vertices and depends on a local neighborhood. The second step traces gradient paths from each $(n-1)$-saddle towards extrema. This tracing requires gradients to be computed only for a small subset of regular vertices, thereby resulting in a low computational load. The memory footprint is also low because the first step requires a simple query within a constant sized local neighborhood and the second step requires only the end point of the path to be stored. 

There exist multiple  parallel algorithms for flood fill. However, these method do not scale due to the communication required to handle the merge events and subsequent serialization of the descending manifold growth computation. In contrast, the first step of the gradient path tracing is amenable to fine grained parallelism, is simple, and highly scalable. The second step can also be accelerated via task parallelism. These advantages motivate us to develop a gradient path tracing algorithm for parallel computation of the extremum graph. In the following, we describe the two steps of the algorithm with a focus on how the steps are parallelized, followed by a discussion on effective simplification of the extremum graph to make it suitable for applications.

\subsection{Grid tessellation}
\label{sec:gridtessellation}

\begin{figure*}[bt]
    \centering
    \begin{tabular}{ccc}
        \subcaptionbox{}{\includegraphics[scale=0.25]{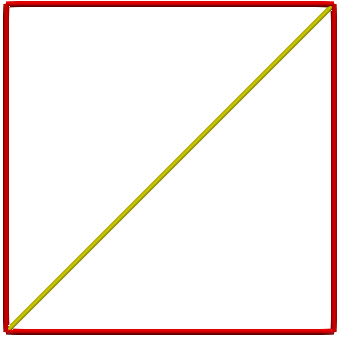}} &
        \subcaptionbox{}{\includegraphics[scale=0.25]{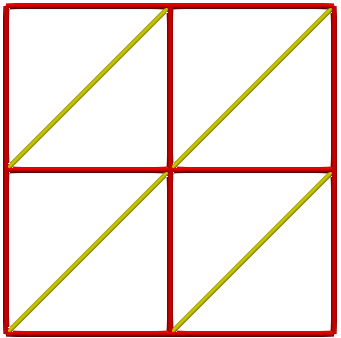}} &
        \subcaptionbox{}{\includegraphics[scale=0.25]{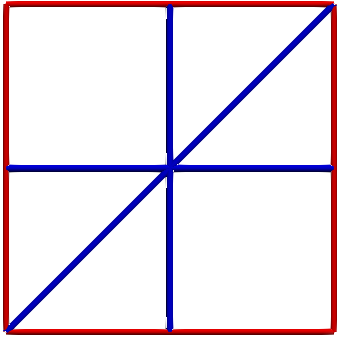}} \\
        \subcaptionbox{}{\includegraphics[scale=0.25]{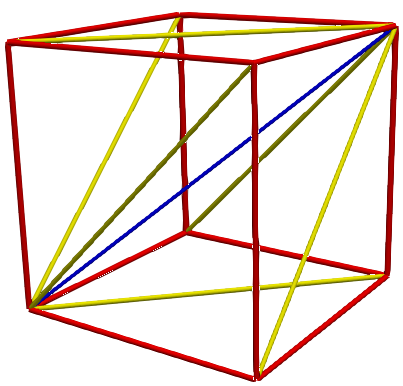}} &
        \subcaptionbox{}{\includegraphics[scale=0.25]{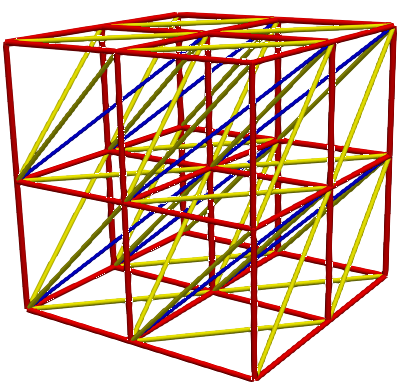}} &
        \subcaptionbox{}{\includegraphics[scale=0.25]{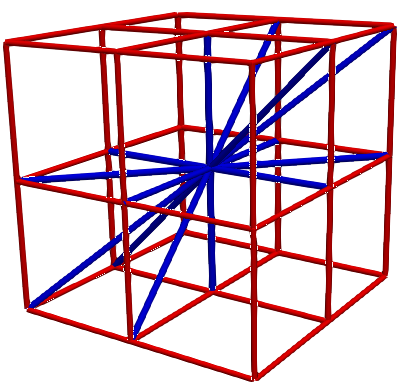}} \\
    \end{tabular}   
    \caption{Tessellating 2D and 3D grids. (a)~A 2D cell is decomposed into two triangles. (b)~Tessellating a 2D grid. (d)~A 3D grid cell is decomposed into six tetrahedra. (e)~Tessellating a 3D grid. The decomposition of the common faces between two adjacent grid cells is consistent. (c,f)~Edges of the neighborhood graph (blue) of the vertex at the centre. The vertex has 6 neighbors in a 2D grid and 14 neighbors in a 3D grid.}
    \label{fig:tessellation}
\end{figure*}
 

We assume that the scalar field is available as a collection of samples at vertices of an $n$-dimensional grid.  Vertices of the grid that represent the domain $\mathcal{D}^n$ have integral coordinates, 
$$V(\mathcal{D}^n) = \{ v \mid v \in \underbrace{\mathbb{Z} \times \mathbb{Z} \times \cdots  \times \mathbb{Z}}_{n} \}. $$
Edges of the grid are between two vertices that differ in exactly one coordinate by a value of 1. Methods for topological analysis and visualization of this scalar field  \highlightadd{typically assume an input piecewise multilinear or piecewise linear scalar function. The samples at vertices may be extended into a scalar function} by either using a multilinear interpolant (trilinear in 3D) within each grid cell or by tessellating the grid cells into linear cells (tetrahedra in 3D) followed by linear interpolation within each linear cell. In either case, the local maxima of the scalar field are located at grid vertices. 

We tessellate the uniform grid into irregular linear cells (tetrahedra in 3D) with the aim of simplifying the extremum graph computation and to support future extensions of the method to unstructured grids. We choose a tessellation that decomposes all grid cells in the same manner and one that is consistent on the common face between adjacent cells, \highlightadd{called the Freudenthal subdivision~\cite{carr2006artifacts}}. Figure~\ref{fig:tessellation} shows the tessellated grid in 2D and 3D (for illustration). The resulting edge set of the tessellated grid $E(\mathcal{D}^n)$ can be stored implicitly and recovered using a simple routine, see Algorithm~\ref{alg:point_adjacency}. The algorithm essentially dictates that two distinct vertices are connected by an edge if and only if the difference vector between the two points consists entirely  of non-negative or non-positive values and the magnitude of the non-zero values is 1. 

The edge set of the tessellated grid
$$ E(\mathcal{D}^n) =  \{ (u, v) \mid u,v \in V(\mathcal{D}^n) \wedge \textnormal{\textit{GridAdjacency}}(u, v) \}. $$
%
%
Figure~\ref{fig:tessellation} also shows the edges incident on a given vertex of the tessellated grid. It follows from the symmetry of the edge set that the number of edges incident on a vertex $v$ is equal to twice the number of non-zero difference vectors, namely $2 \times (2^n-1)$ for nD grids. So, the number of vertices in the neighborhood of $v$ in a 3D grid equals 14. Critical points are identified and classified based on a connected component labeling of this neighborhood. We discuss this classification next.  

\begin{algorithm}
  \SetAlgoLined
  \LinesNumbered
  \caption{\textit{GridAdjacency}}
  \label{alg:point_adjacency}
  \KwIn{ $p, q:$ Two vertices. Test if they are adjacent in the tessellated $n$-dimensional grid.}
  \KwResult{$True$ if $p,q$ are adjacent and $False$ otherwise}

  $n \leftarrow len(p)$ \hfill \text{\color{blue} $\triangleright$ $p$ is a vertex in an $n$-dimensional grid}\\
  $U \leftarrow \phi$

  \ForEach{$i \in 1..n$}{
      $d \leftarrow p_i - q_i$ \hfill \text{\color{blue} $\triangleright$ difference between $i^{th}$ coordinates}\\
      $U \leftarrow U \cup \{d\}$ \hfill \text{\color{blue} $\triangleright$ collect component wise difference}\\
  }
%
%
  \If{$U \highlightadd{\subseteq} \{0, 1\}$ or $U \highlightadd{\subseteq} \{0, -1\}$}
  {
      \Return {$True$}
  }
  \Else
  {
      \Return {$False$}
  }

\end{algorithm}

 \subsection{Critical point classification}
 \label{sec:pointclassification}

Critical points of the piecewise linear function defined over a tessellated grid can be identified and classified based on a local neighborhood~\cite{Ban70,edelsbrunner2003morse}. All critical points are located at vertices of the grid. The \emph{star} of a vertex $v$ consists of $v$ together with the collection of edges, triangles, tetrahedra, and higher dimensional cells incident on it. Figure~\ref{fig:tessellation}(f) shows the edges (blue) of the star of a vertex in a cube grid. The collection of end points of these edges, excluding $v$, together with the induced edges and triangles form the \emph{link} of $v$. The star and link are two useful notions of neighborhood of a vertex within the tessellated grid. Assuming that there are no degeneracies, scalar values at vertices in the link are either lower or higher than at $v$. Vertices of the link with scalar value lower than at $v$ together with the induced edges and triangles form the \emph{lower link} of $v$. Similarly, the \emph{upper link} of $v$ is defined as the collection of vertices of the link with scalar values greater than at $v$ together with the induced edges and triangles. 

Critical points are classified based on the number of connected components of the upper and lower link, denoted $\beta_0^+$, $\beta_0^-$ and called the zeroth Betti number~\cite{edelsbrunner2003morse}. Figure~\ref{fig:vertex-classification} illustrates the upper (red) and lower (green) link for various types of critical points and for a regular point. Since our objective is to compute the extremum graph, we are specifically interested in identifying the maxima and 2-saddles. The first step of the algorithm visits each grid vertex and classifies it as critical or regular as shown in Table~\ref{tab:vertex-classification}. 
\begin{table}[h]
\centering
\begin{tabular}{r|rr}
		    & $\beta_0^+$	& $\beta_0^-$ \\ \hline \\
maximum	    & 0			    & 1 		\\	 
$(n-1)$-saddle	& $\geq$2	    & $\bullet$ \\
1-saddle	& $\bullet$	    & $\geq$2	\\
minimum	    & 1			    & 0		\\
regular	    & 1			    & 1
\end{tabular}   
  \caption{Classifying a vertex based on the topology of the upper and lower link. $\beta_0^+$ and $\beta_0^-$ count the number of connected components of the upper and lower link, respectively. $\bullet$ indicates that the value does not affect the classification.}
  \label{tab:vertex-classification}
\end{table}

\newcommand{\critPtFigsScale}{0.12}

\begin{figure*}[bt]
    \centering
    \begin{tabular}{ccc}

    \subcaptionbox{Maximum}{\includegraphics[scale=\critPtFigsScale]{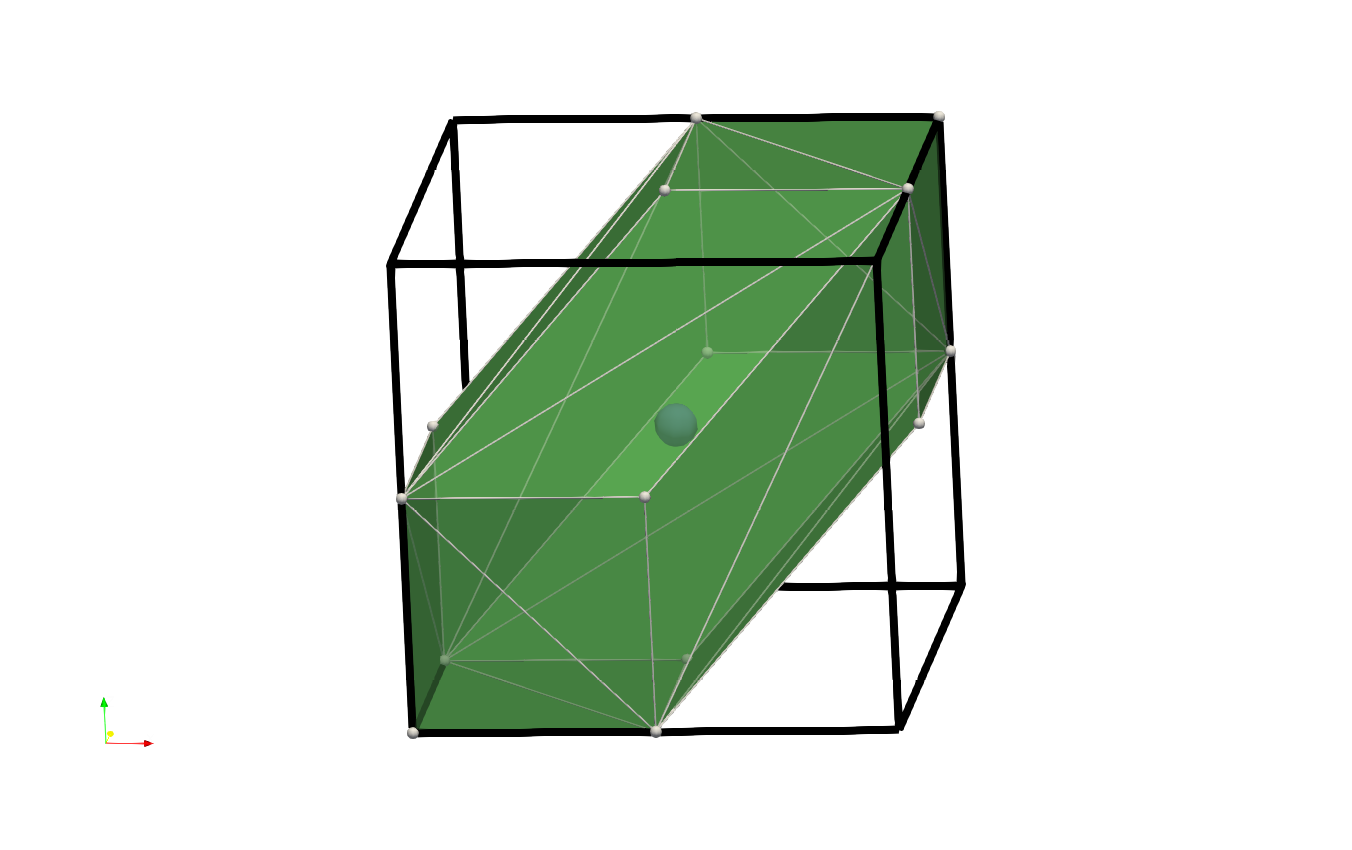}} &
    \subcaptionbox{2-Saddle}{\includegraphics[scale=\critPtFigsScale]{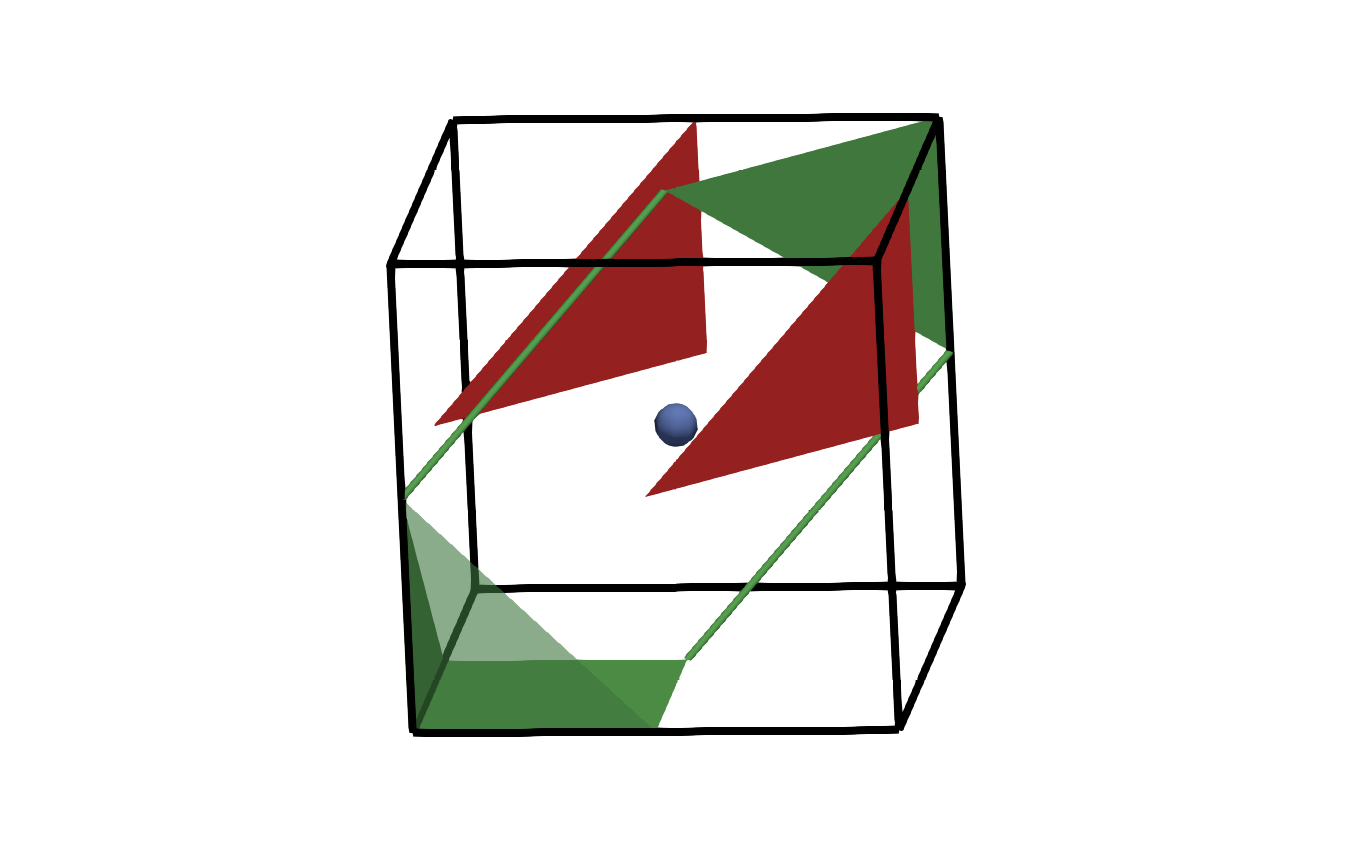}} &
    \subcaptionbox{Regular point}{\includegraphics[scale=\critPtFigsScale]{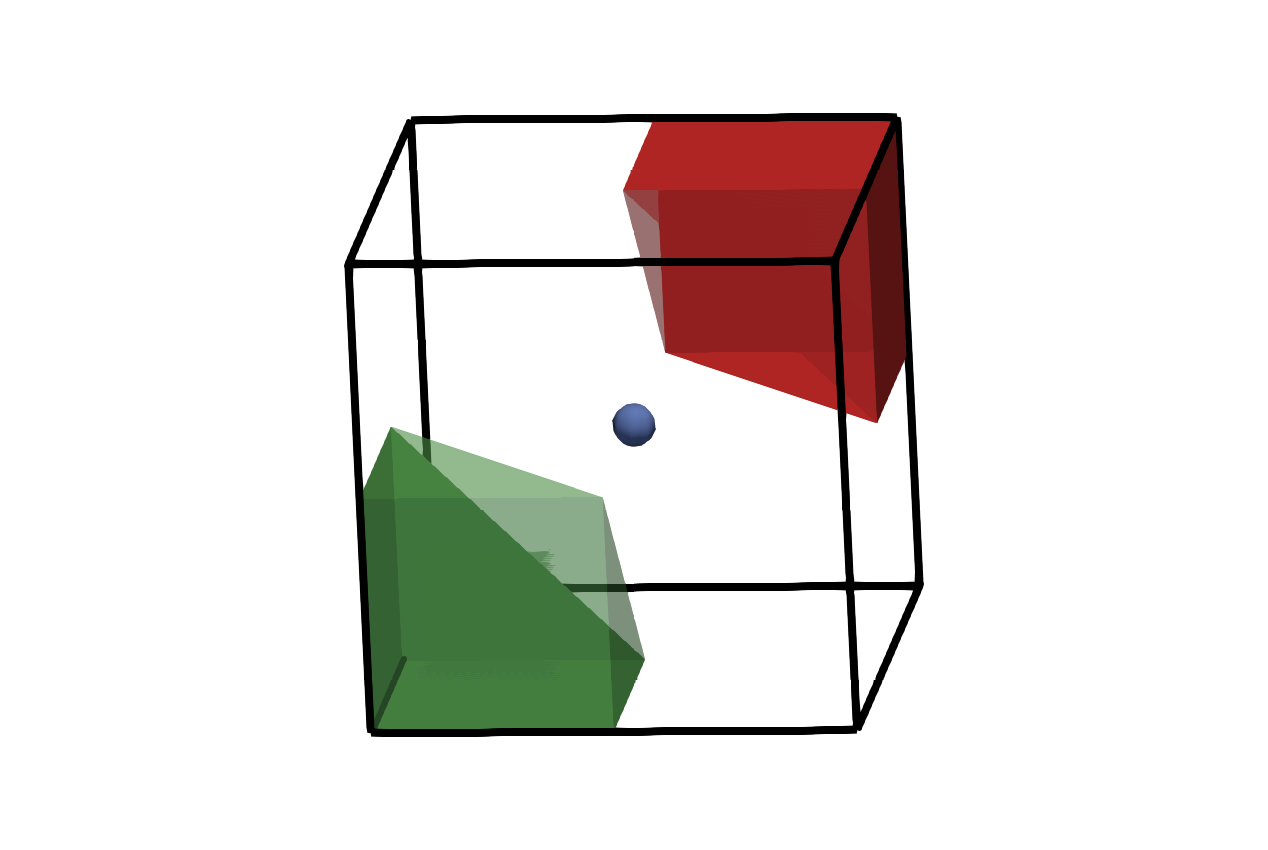}} \\

    \subcaptionbox{Minimum}{\includegraphics[scale=\critPtFigsScale]{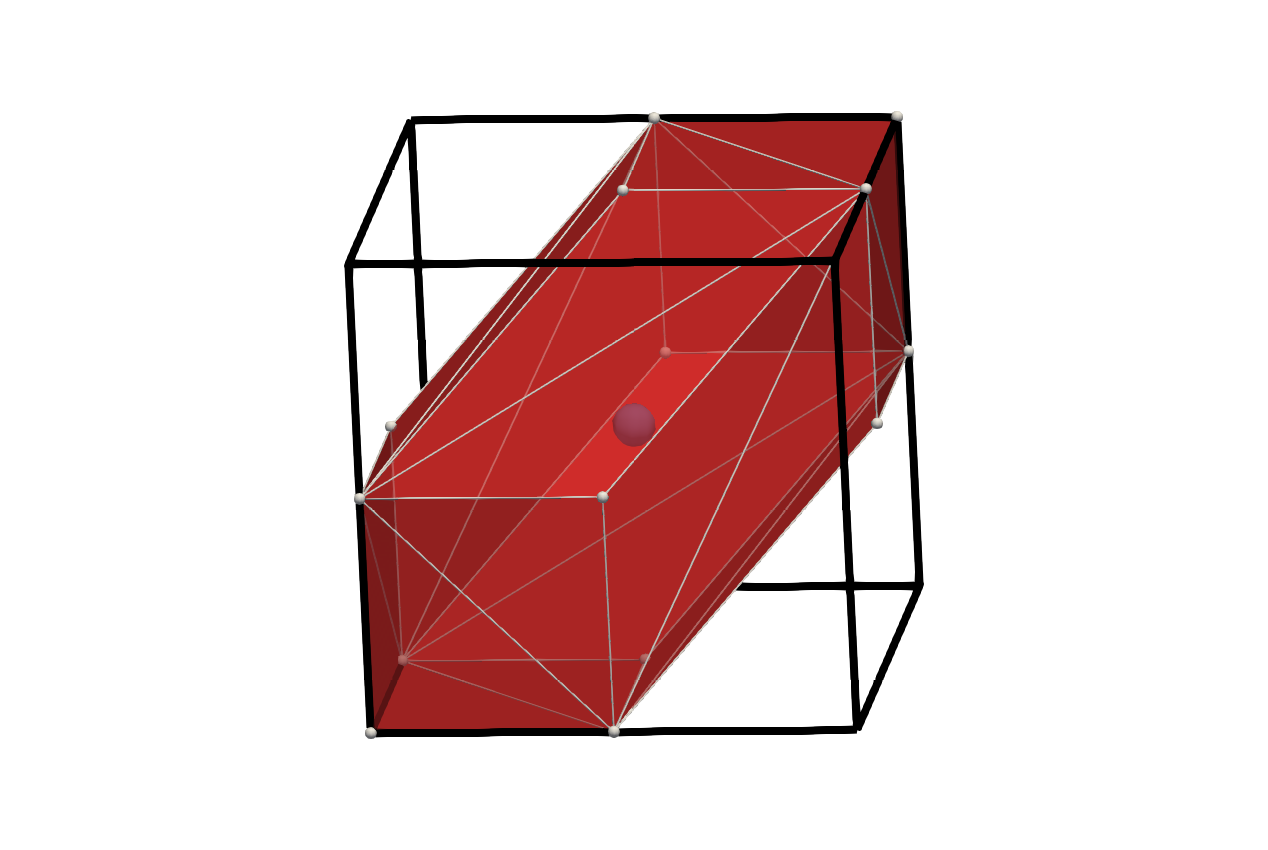}} & 
    \subcaptionbox{1-Saddle}{\includegraphics[scale=\critPtFigsScale]{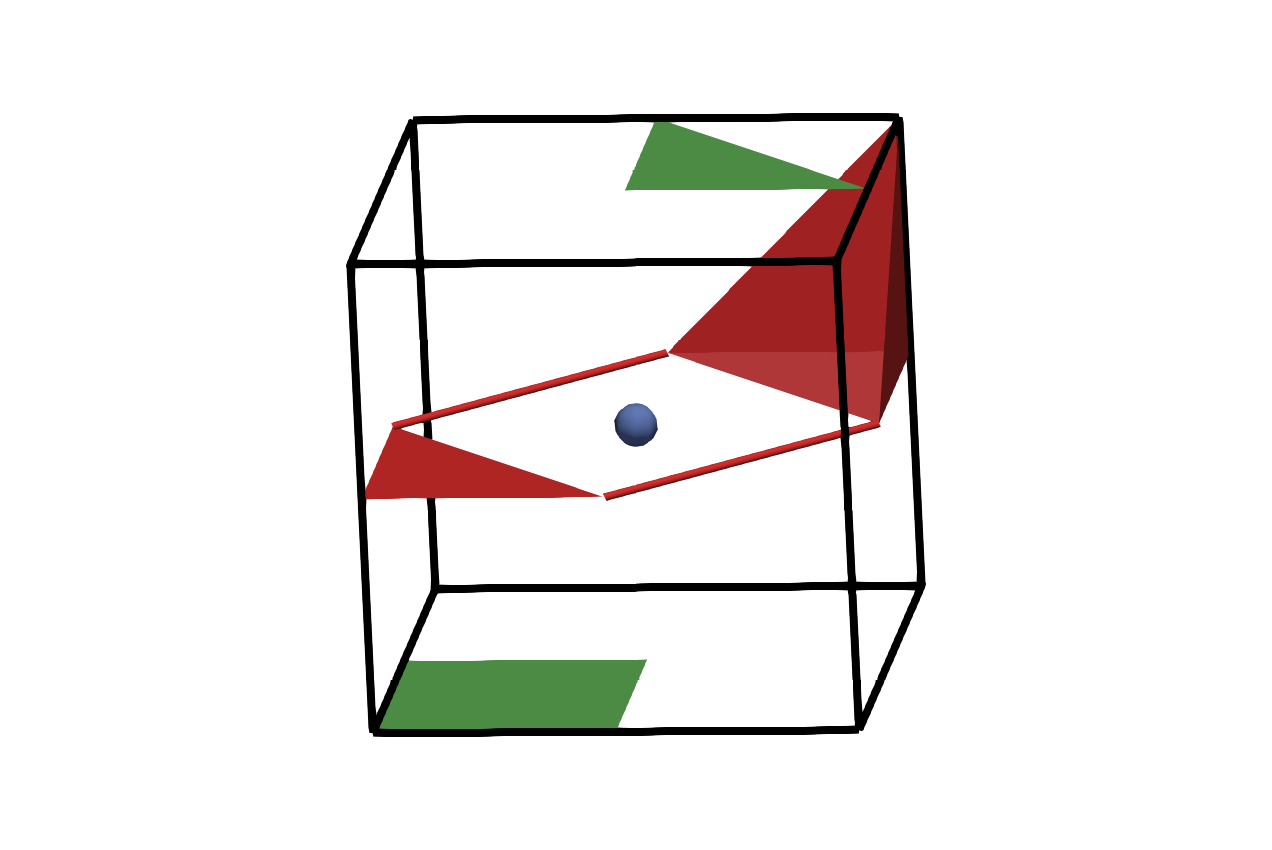}} &
    \subcaptionbox{Multi-saddle}{\includegraphics[scale=\critPtFigsScale]{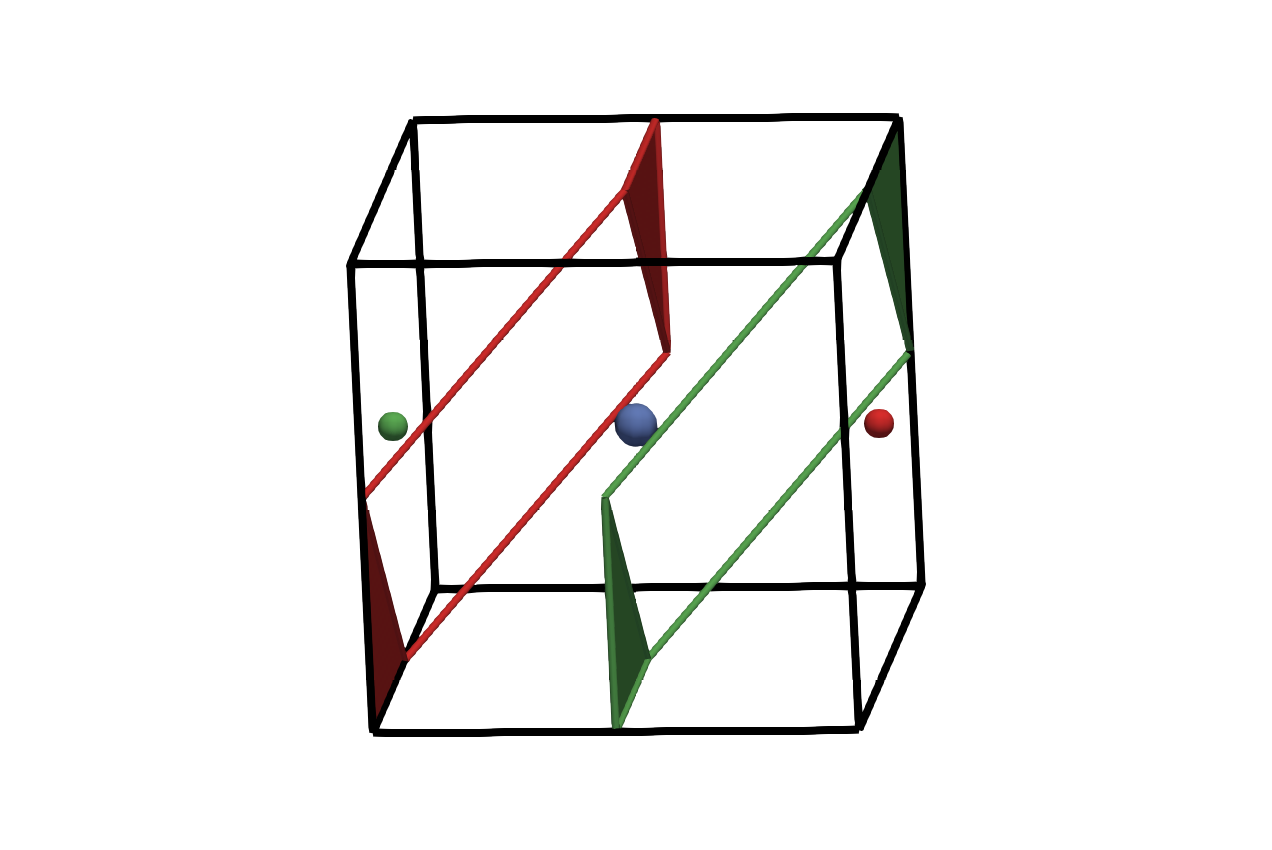}}

    \end{tabular}
    \caption{Critical points can be identified and classified based on the upper (red) and lower (green) link connectivity. \highlightadd{The upper link of a maximum~(a) is empty and the lower link is equivalent to a sphere, whereas the upper link of a minimum~(d) is equivalent to a sphere and the lower link is empty. The upper link of a simple 2-saddle~(b) consists of two connected components and its lower link consists of a single component, whereas the upper link of a 1-saddle~(e) consists of a single component and the lower link consists of two connected components. Both the upper and lower links of a regular point~(c) consist of a single component.  The multi-saddle~(f) is a degenerate structure that does exist in piecewise linear functions.}}
    \label{fig:vertex-classification}
\end{figure*}


Both $\beta_0^+$ and $\beta_0^-$ are typically computed by performing a BFS graph traversal on the upper and lower link of the vertex $v$. The BFS traversal of the upper and lower link is an embarrassingly parallel task across all vertices. We utilize GPU parallelism to perform this traversal in a massively parallel fashion by launching a CUDA thread for each vertex. Our implementation \highlightadd{moves this task automatically} to a multicore CPU in the absence of a CUDA enabled GPU.

\highlightadd{In order to make the computation GPU friendly, we utilize a union-find based connected component tracking instead of the BFS traversal, similar to previous approaches~\cite{carr2019scalable}}. The union-find data structure is initialized with a collection of singleton sets, each containing a vertex of the link of $v$. Next, we test for the existence of an edge between a pair of vertices in the link using Algorithm~\ref{alg:point_adjacency}. If the edge exists and both end points of the edge belong to the upper link (lower link), we perform a union operation on the corresponding sets. Degeneracies are handled using a simulated perturbation~\cite[Section 1.4]{EH09}, which consistently determines if the scalar value at a link vertex is lower or higher than the value at $v$.

After identifying the connected components in the link, we label each component as upper or lower link by testing one vertex within the component and hence compute $\beta_0^+$ and $\beta_0^-$. This method is GPU friendly as all working threads test the same number of edges and require almost the same time, thereby causing a very low thread divergence. If a vertex is identified as regular, we additionally compute the gradient of the scalar field at the vertex. The gradient is approximated as the vector towards the vertex with the highest scalar value in the upper link.

\subsection{Path tracing}
\label{sec:pathtracing}
We opt to use a direct path tracing algorithm to trace the saddle-maximum arcs that constitute the extremum graph over the flood-fill approach. \highlightadd{The primary reason for this choice is the benefit in terms of scalability of the computation. In addition,} path tracing does not require each vertex of the grid be visited. Further, a flood-fill based algorithm requires an ordered list of vertices, sorted on the scalar values. 

\begin{algorithm}
    \SetAlgoLined
    \LinesNumbered
    \caption{\textit{TraceGradientPaths}}
    \label{alg:path_tracing}
    \KwIn{ $s, M$: Source $(n-1)$ saddle and the set of all maxima}
    \KwResult{Set of gradient paths originating from $s$}
    $P \leftarrow \phi$

    \ForEach{$u \in \textit{UpperLinkRep}(s)$}{

        $p \leftarrow s$\hfill \text{\color{blue} $\triangleright$ Initialize path $p$ with $s$}\\

        \While{$u \not\in M$}{
            $p \leftarrow p \,\, || \,\, u$ \hfill \text{\color{blue} $\triangleright$ Append $u$ to $p$}\\
            $u \leftarrow \textit{gradient}(u)$\hfill \text{\color{blue} $\triangleright$ Follow gradient at $u$}\\
        }

        $P \leftarrow P \,\, \cup \,\, \{p\}$
    }

    \Return{$P$}

\end{algorithm}

The path tracing algorithm requires only the gradient at vertices. The gradient paths in an extremum graph originate from $(n-1)$-saddles, so the algorithm visits only a small fraction of vertices of the grid. However, this approach is not without challenges. The operation of tracing a gradient path is inherently serial in nature. One approach towards parallelism is to trace the gradient paths for all $(n-1)$-saddles concurrently.
%

Algorithm~\ref{alg:path_tracing} computes the collection of gradient paths that originate at an $(n-1)$-saddle. The subroutine \textit{UpperLinkRep} returns the highest vertex in each upper link component of the saddle. For each such vertex, it iteratively follows the gradient vectors until termination at a maximum. This algorithm is executed in parallel for each $(n-1)$-saddle. Unlike the critical point classification, Algorithm~\ref{alg:path_tracing} is not amenable to efficient GPU parallelism. Varying lengths of the gradient paths causes unequal division of work among CUDA threads, and results in significant thread divergence.

\begin{figure}
    \centering
    \includegraphics[scale=0.12]{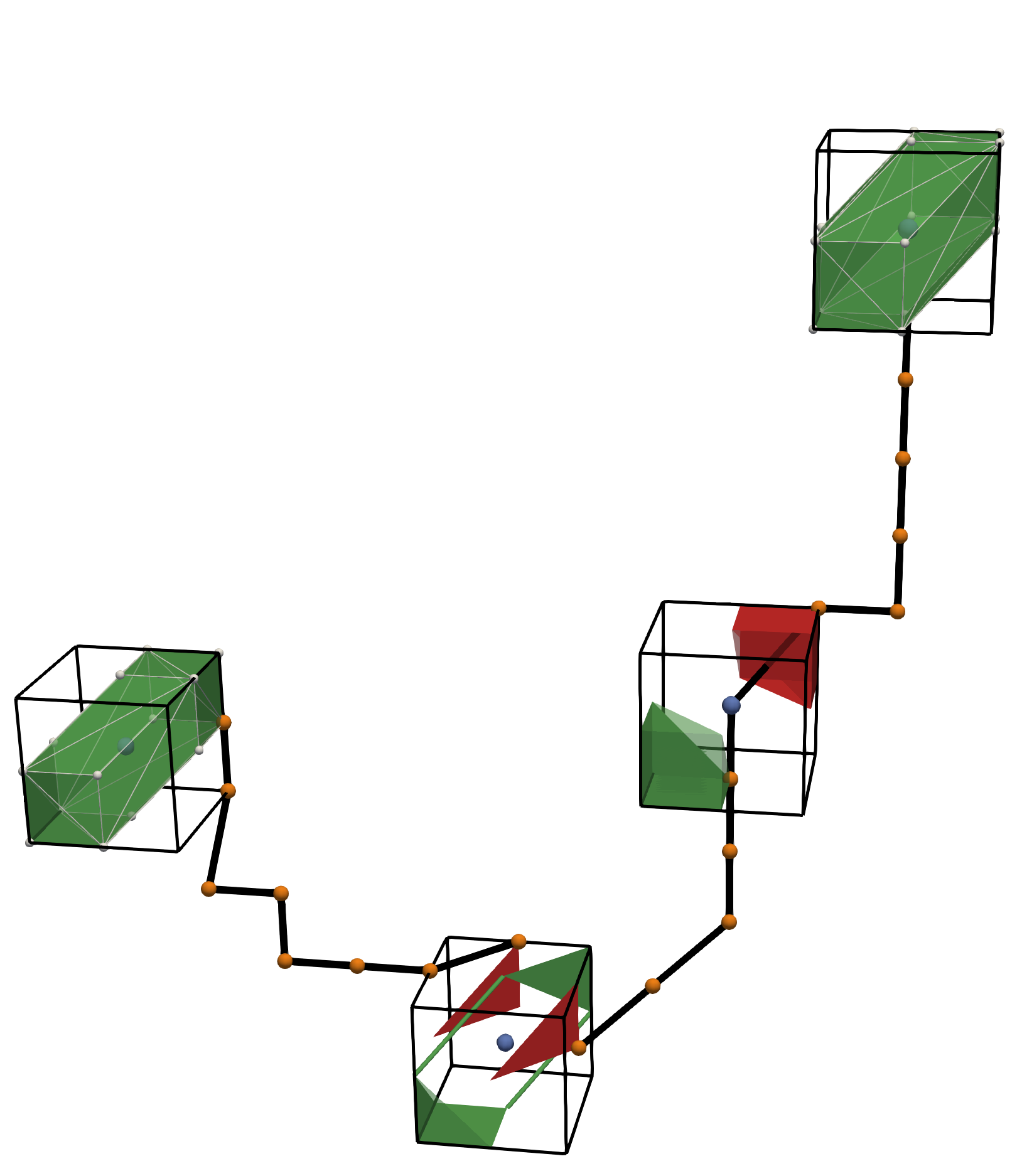}
    \caption{Gradient path tracing for computing arcs of extremum graph. Ascending 1-dimensional manifolds are computed for each 2-saddle. The gradient paths that constitute these ascending manifolds represent the arcs of the extremum graph. Note that they pass through regular points, entering and exiting its neighborhood via the lower and upper link, respectively.}
    \label{fig:path_tracing_sample}
\end{figure}

Figure \ref{fig:path_tracing_sample} illustrates the ascending 1-dimensional manifolds for a 2-saddle. Notice how the gradient paths originate from the 2-saddle via each upper link component and terminate at maxima. Given two adjacent regular points (orange) $(p, q)$ on a gradient path, $p$ lies in the lower link of $q$ and $q$ lies in the upper link of $p$. The neighborhood of one regular point is shown in detail. The gradient path enters through its lower link and  exits via its upper link.

\subsection{Extremum graph simplification}
\label{sec:simplification}

\begin{figure*}[!ht]
    \centering
    \begin{tabular}{ccc}
        \subcaptionbox{Fuel}{\includegraphics[scale=0.09858059701492539]{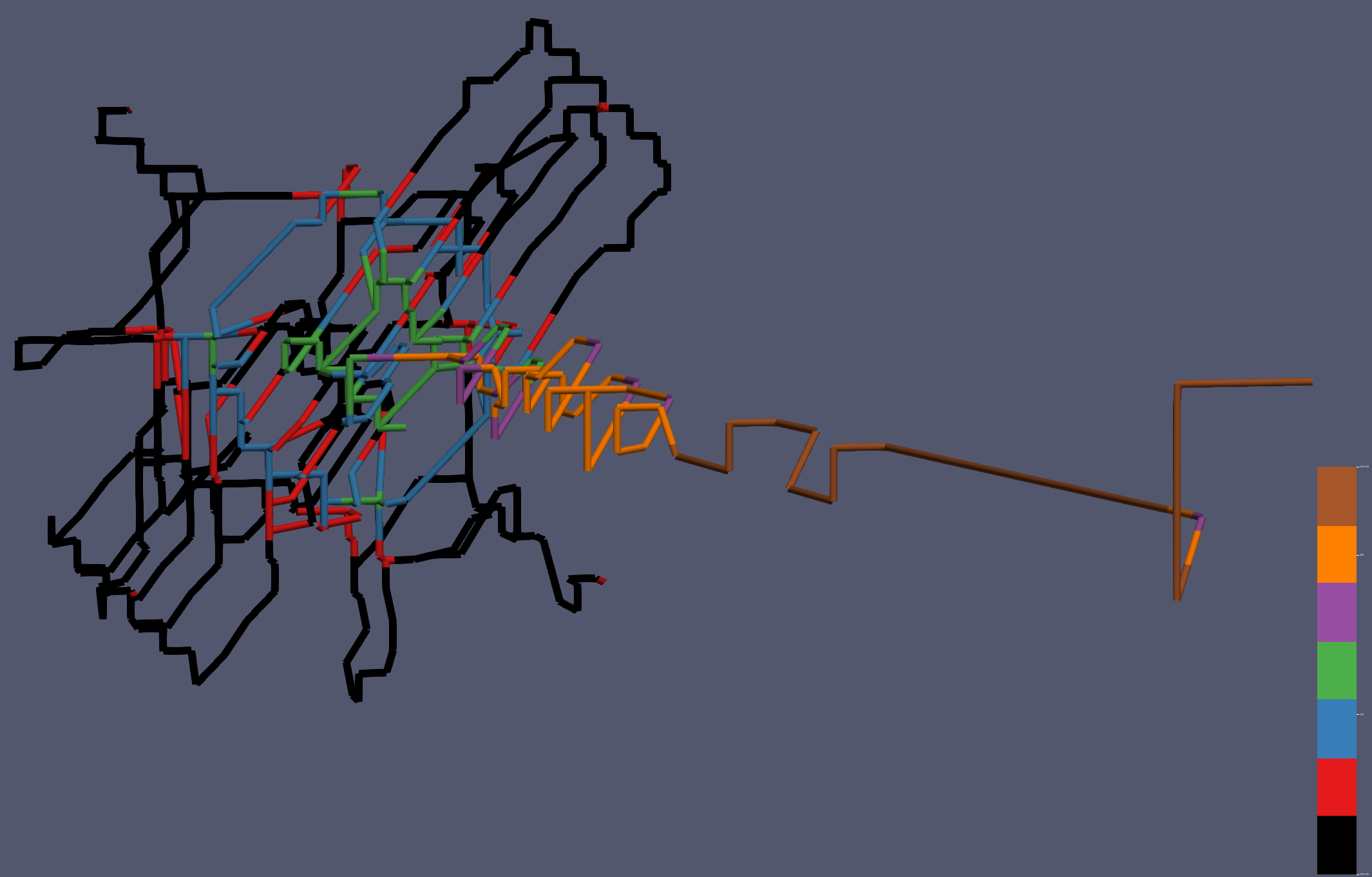}} &
        \subcaptionbox{Lobster}{\includegraphics[scale=0.08630597014925373]{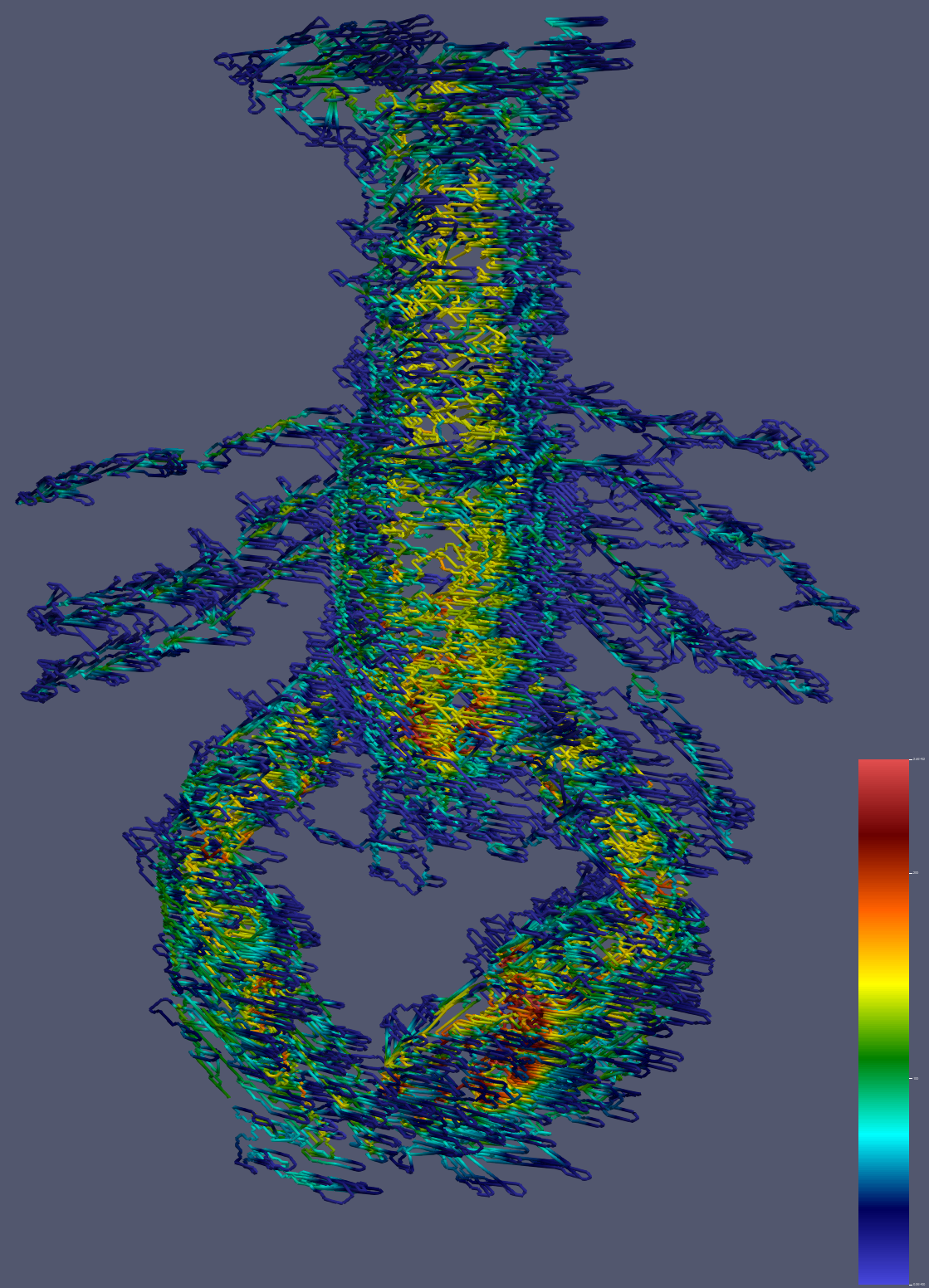}} &
        \subcaptionbox{Foot}{\includegraphics[scale=0.07000373134328358]{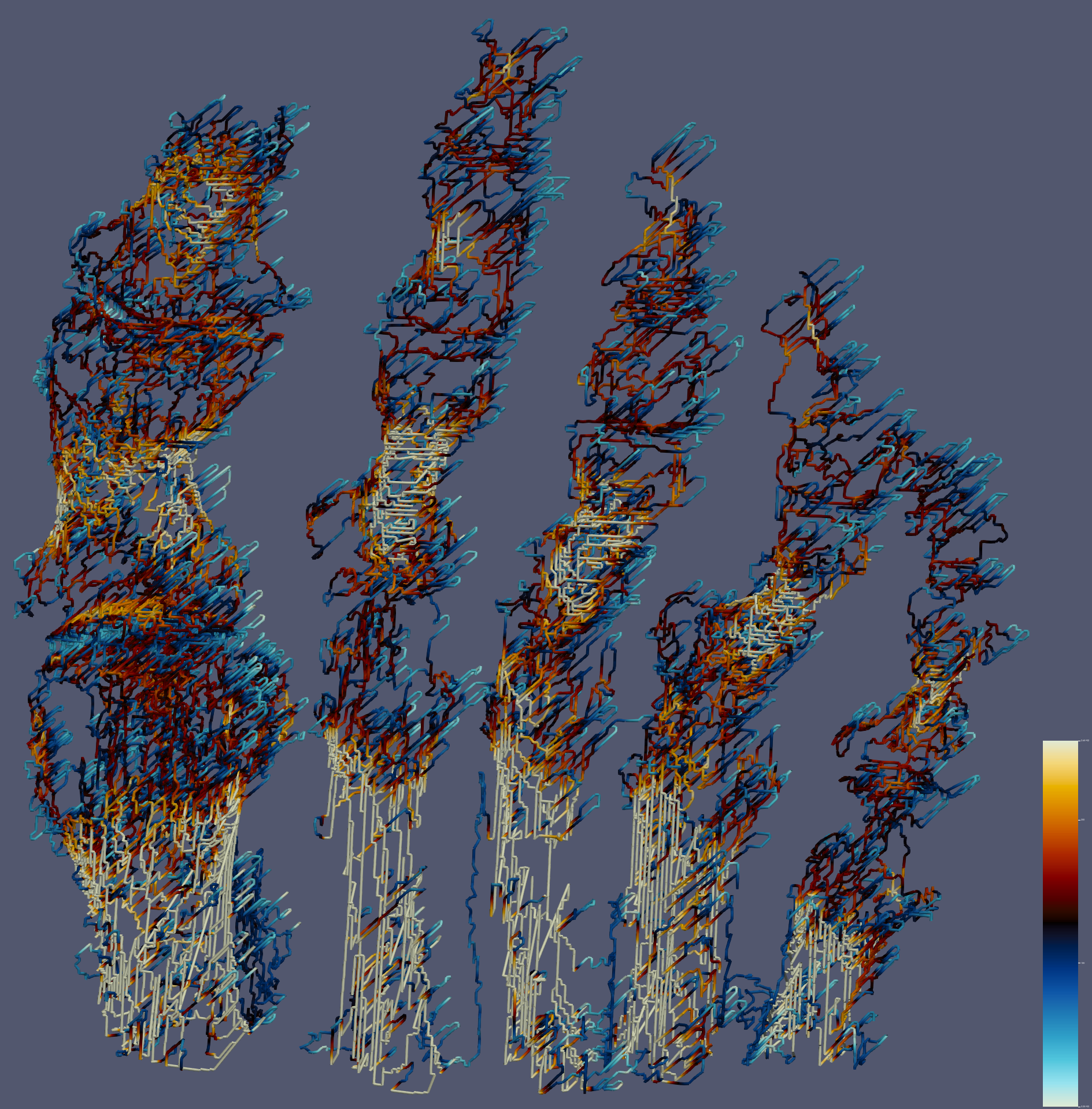}} \\

        \subcaptionbox{Boston Teapot}{\includegraphics[scale=0.11478650826929827,]{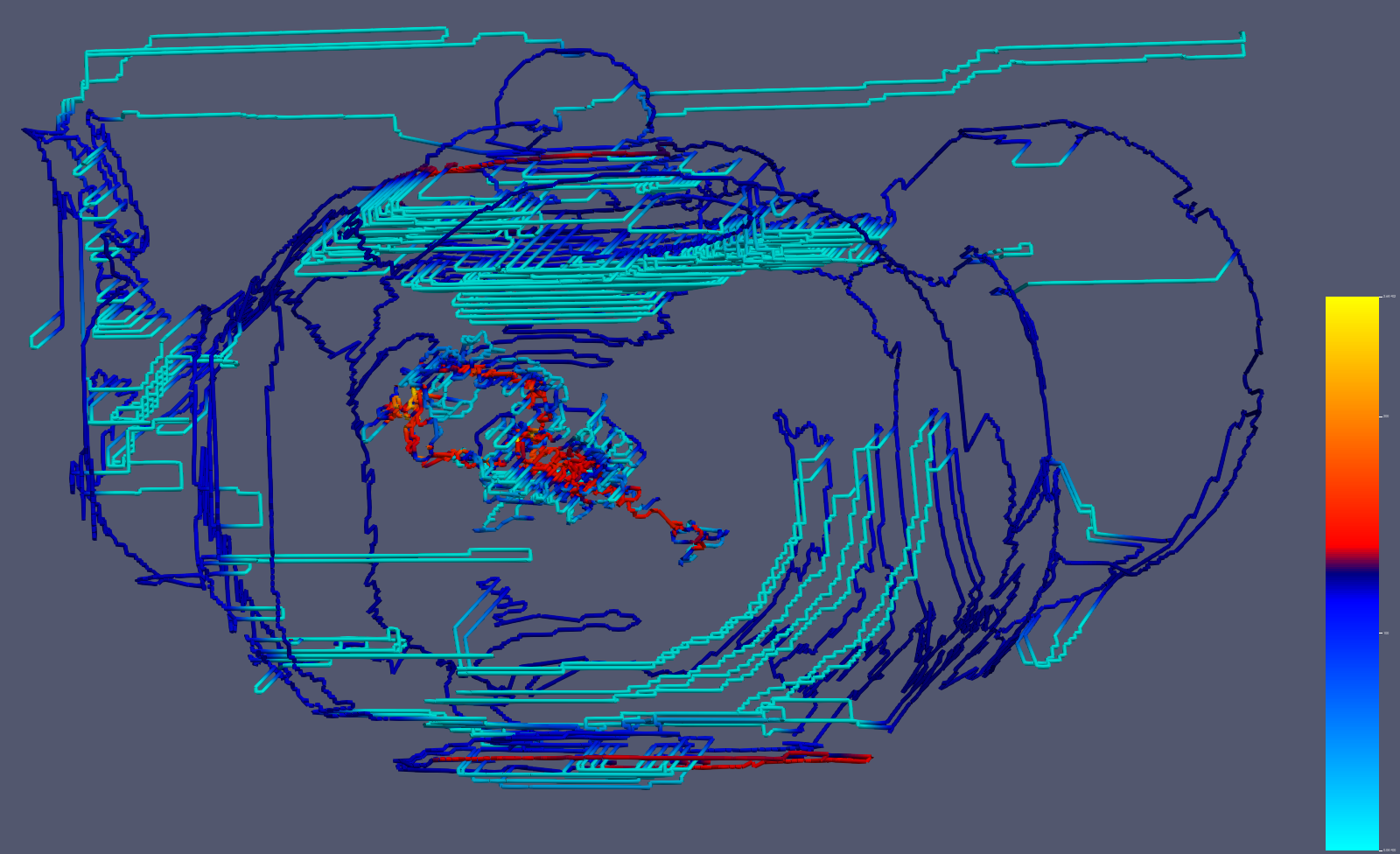}} &
        \subcaptionbox{Nucleon}{\includegraphics[scale=0.06198471446542107,]{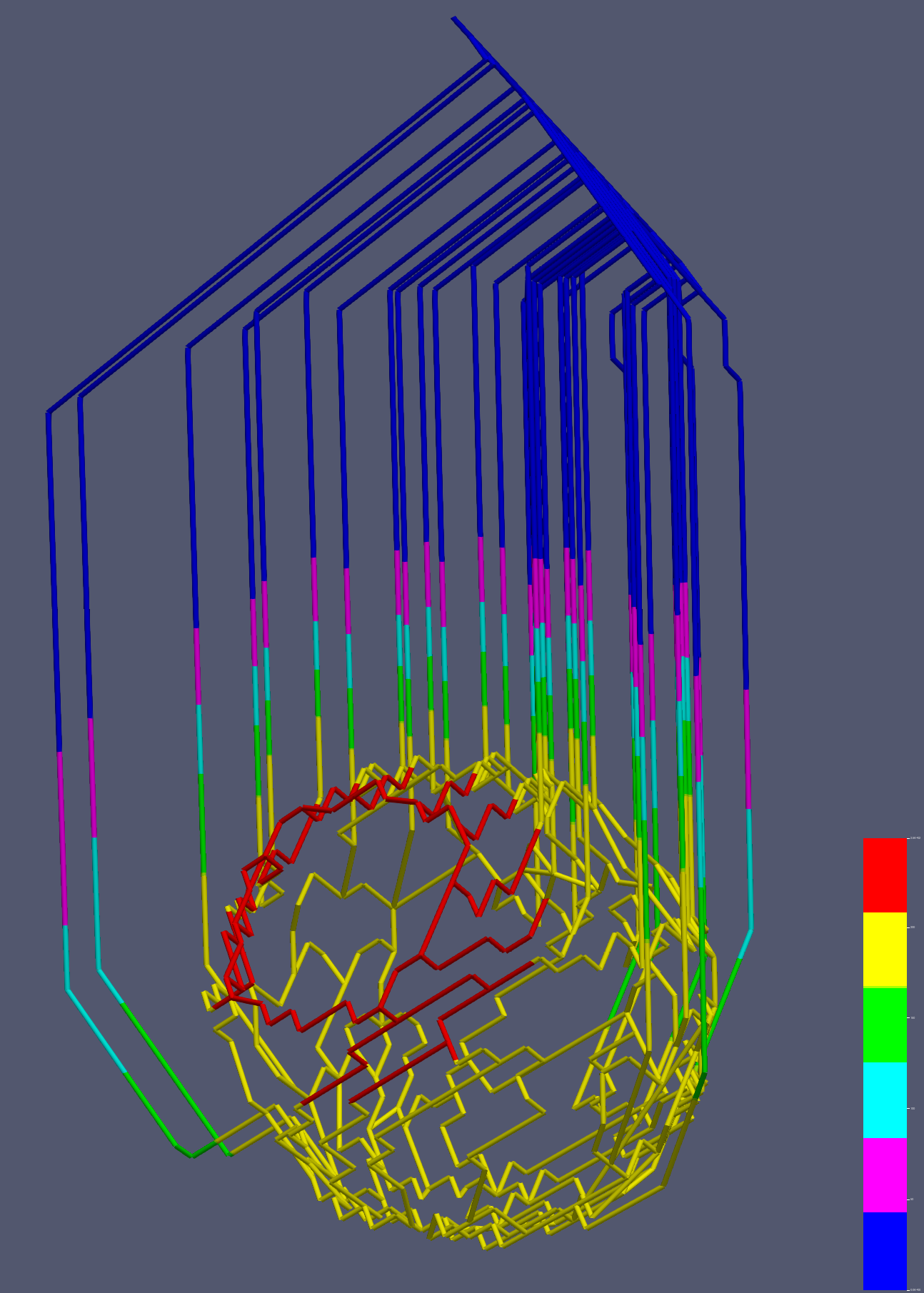}} &
        \subcaptionbox{Neghip}{\includegraphics[scale=0.06173874337627258]{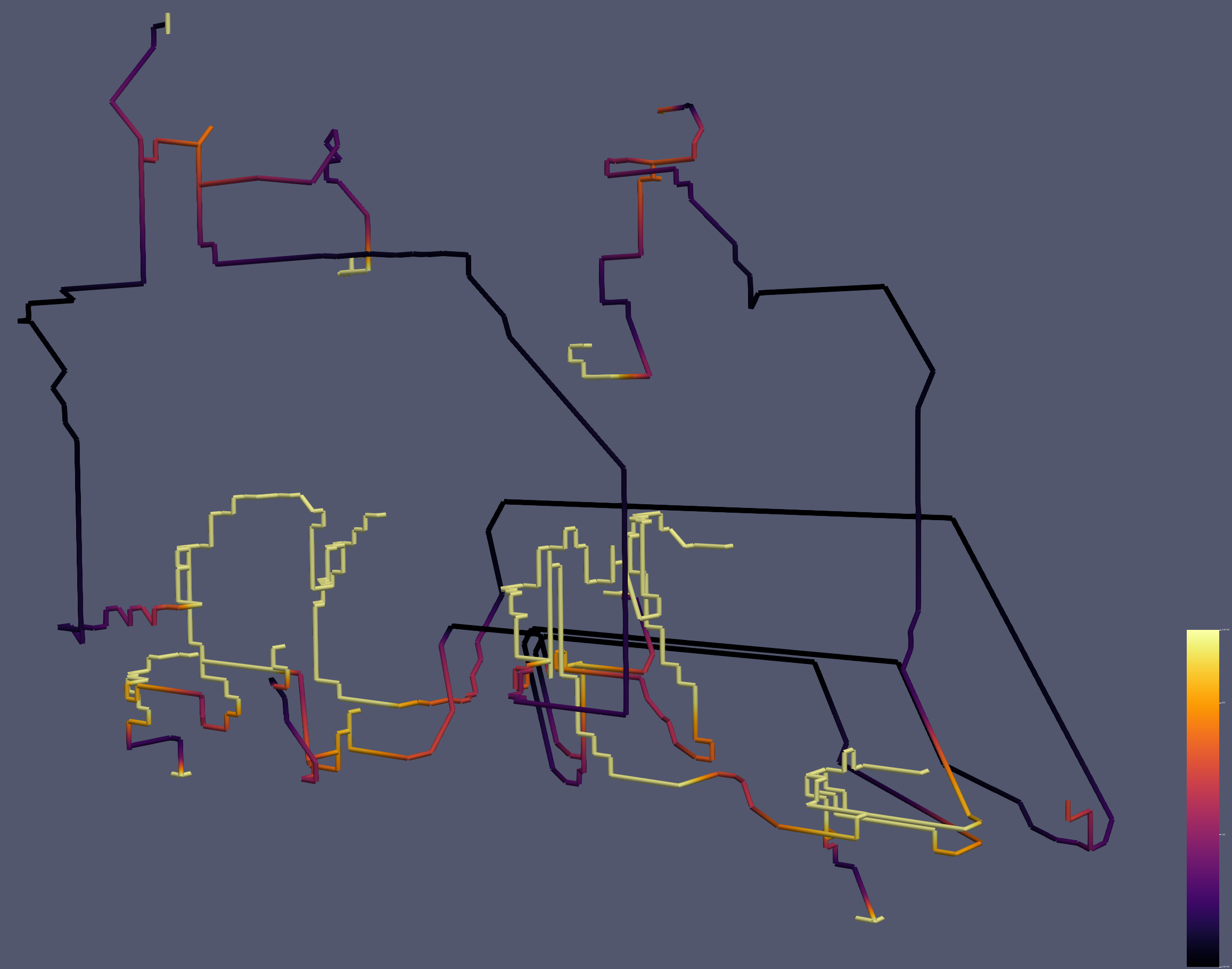}} \\

        \subcaptionbox{Hydrogen}{\includegraphics[scale=0.13541275008182635,]{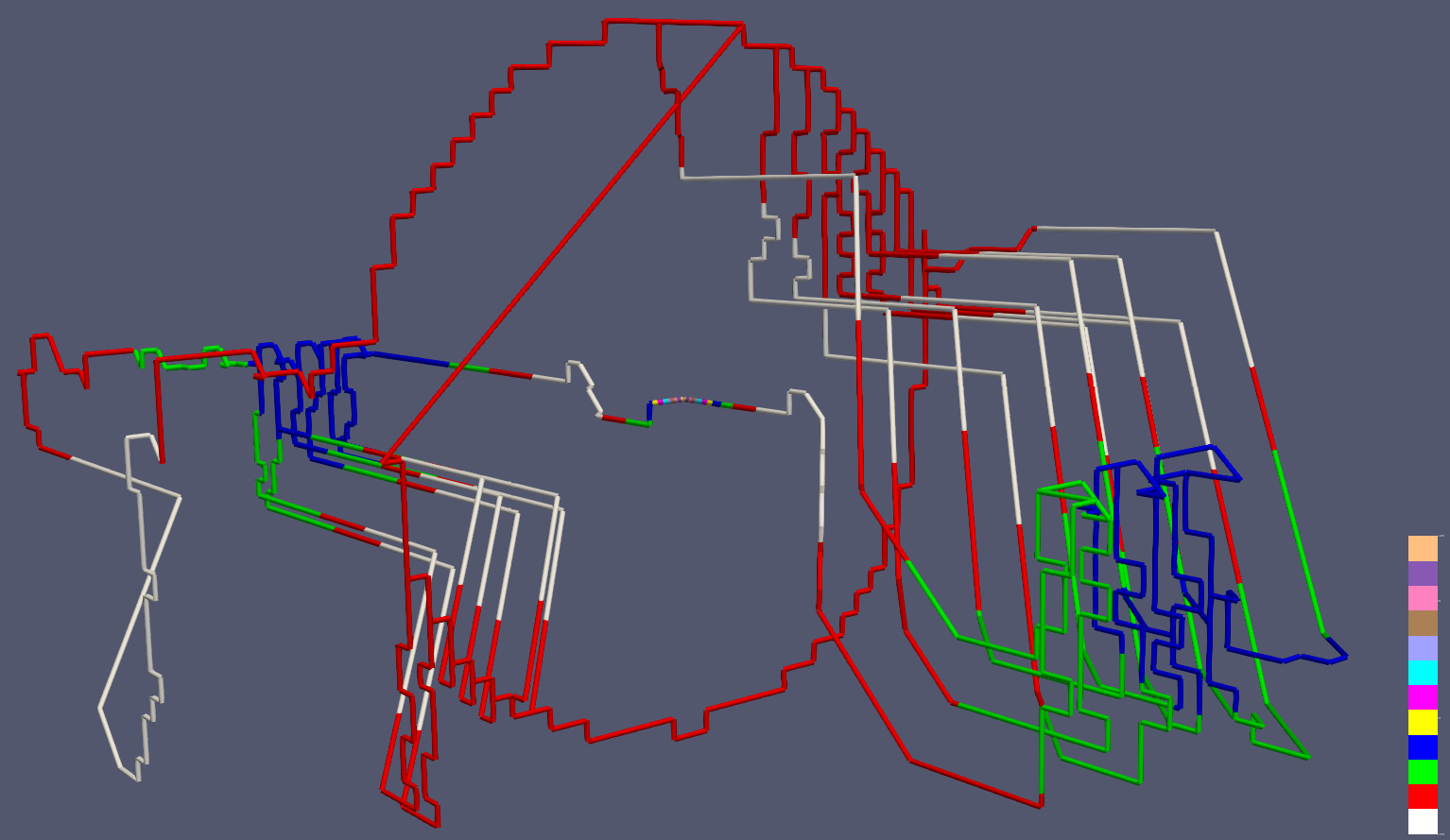}} &
        \subcaptionbox{Engine}{\includegraphics[scale=0.08105759706652484,]{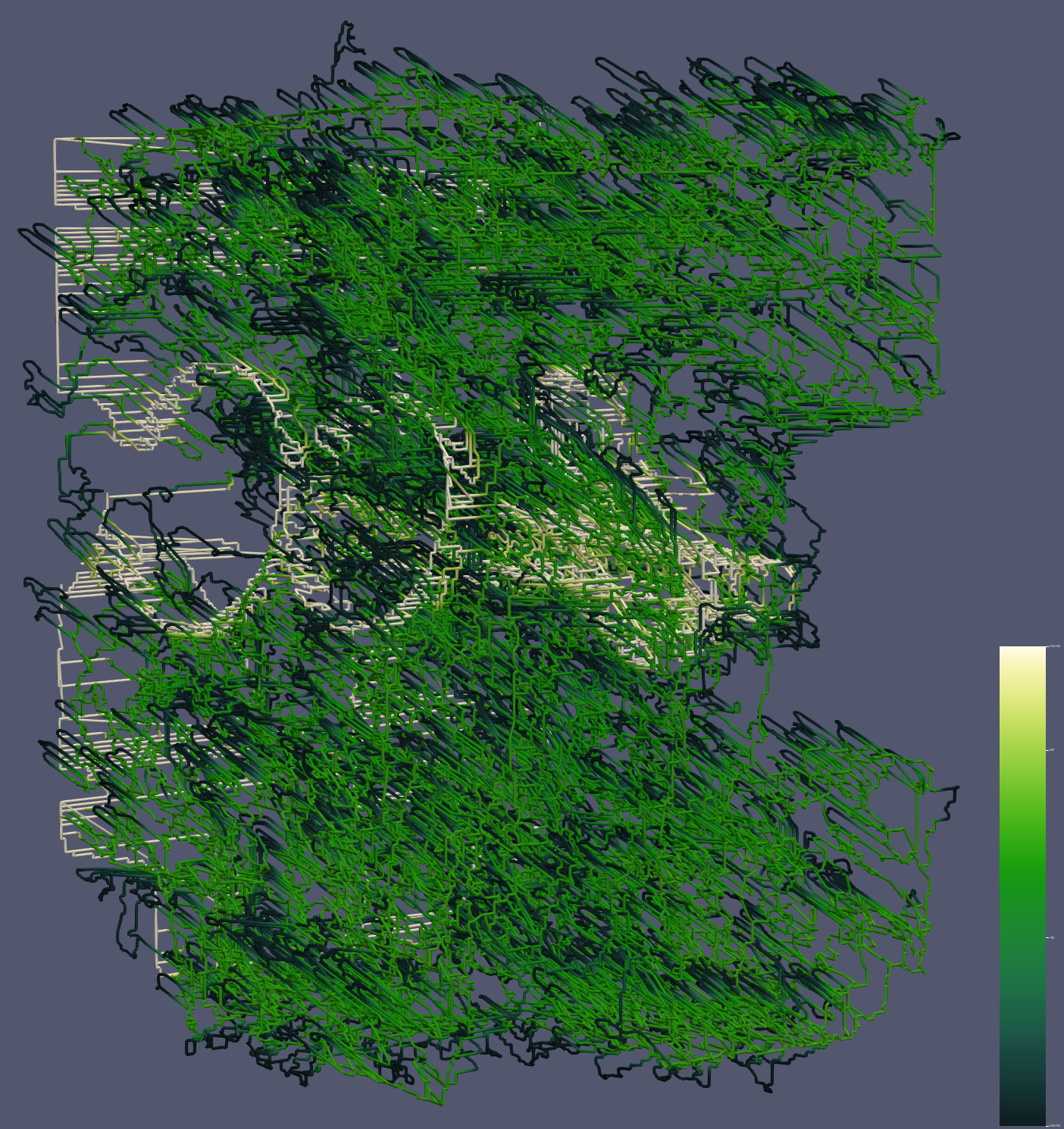}} &
        \subcaptionbox{Heptane Gas}{\includegraphics[scale=0.06281251073971034]{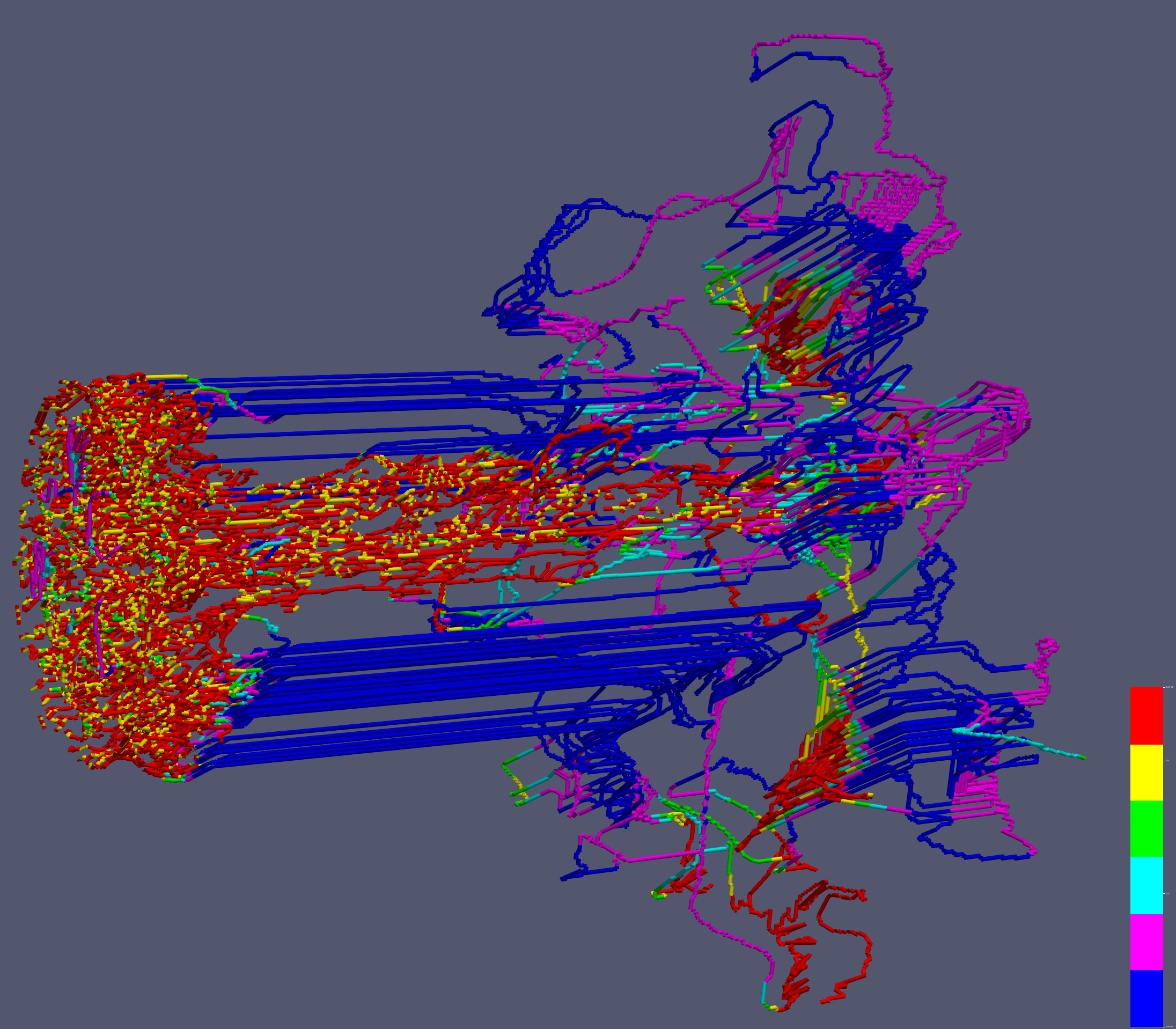}} \\



    \end{tabular}
    \caption{Simplified extremum graphs for various datasets. Nodes (extrema and saddles) are not displayed to reduce clutter. The scalar field is mapped to color as indicated by the legend.}
    \label{fig:simplified-extremum-graph}
\end{figure*}

Scientific data is often noisy, which manifests as an increased number of critical points and hence a larger sized extremum graph. Topological simplification identifies noisy topological features and removes them in a controlled manner, often as a sequence of critical point pair cancellation operations. The simplification is typically directed by the notion of persistence~\cite{ELZ02}.  
Indeed, the extremum graph computed using the algorithm that we describe in the previous sections \highlightadd{contains} a large number of saddles and extrema for all datasets. Several critical points and their incident gradient paths correspond to noisy topological features. The presence of these noisy elements cause occlusion, thereby adversely affecting data visualization tasks, and hinder feature selection. We propose three simplification operations to aid the proper identification of key features from the extremum graph. \highlightadd{The \textit{persistence directed cancellation} and \textit{saturated persistence directed simplification} algorithms are implemented in \parext as they are useful in removing noise and potentially uninteresting features from the extremum graph~\cite{correa2011topological}. We introduce the \textit{arc-bundling} algorithm as a method to remove multiple connections between a pair of extrema.} Any sequence of these operations may be applied on an extremum graph. We describe the simplification operations below. \highlightadd{Their implementations in \parext are serial in nature.}

\myparagraph{Arc bundling.}
The gradient path tracing algorithm computes individual arcs between an $(n-1)$-saddle-maximum pair independent of each other. So, it is likely that a pair of maxima contain more than one common $(n-1)$-saddle in their neighborhood. We select a single representative saddle based on a specified criterion and discard the remaining saddles that are shared by the two maxima within their neighborhood. We term this pruning operation as arc bundling, as this reduces the number of arcs between an extrema pair. Note that this operation does not disturb the connectivity between extrema and the resulting graph continues to represent the Morse decomposition. In our current implementation, we select the saddle with the highest scalar value as the representative.

\myparagraph{Persistence directed cancellation.}
This operation cancels an $(n-1)$-saddle-maximum pair that is connected by an arc in the extremum graph and reconnects the neighborhood of both nodes.  The critical point pairs are ordered based on their difference in scalar value, motivated by the notion of topological persistence, and scheduled for cancellation. The simplification begins by initializing a priority queue with all $(n-1)$-saddles, where the priority is inversely related to the persistence of the saddle. Computing the topological persistence of all saddles upfront is unnecessary and computationally expensive. We use lazy updates where the true persistence of a saddle is computed when it reaches the top of the queue and is ready for cancellation. We initialize the cost of canceling a simple $(n-1)$-saddle as the smallest difference in scalar value between the saddle and its two adjacent maxima in the extrema graph. This cost is recomputed whenever a saddle is removed from the  top of the queue.

The next step of the simplification iteratively removes the saddle $s$ from the top of the queue, recomputes its cost, and decides whether to discard, reinsert, or cancel the saddle. If the updated cost is above a user-specified  persistence threshold, the saddle should not be cancelled. Note that subsequent cancellations will not decrease the cost of this saddle and so, it is safely discarded. If the updated cost is greater than the cost of the top of the queue (but smaller than the persistence threshold) then the saddle is reinserted into the queue because the queue contains other saddles with lower costs that need to be canceled. Else, the saddle should be canceled. The gradient path from $s$ towards its persistence pair $m$ is reversed to cancel the critical point pair. The above step is repeated until the queue is empty. This  simplification removes all low persistence critical point pairs (arcs) from the extremum graph, while retaining features of potential interest.

Multi-saddles have greater than two arcs incident on them. The cost of a multi-saddle is computed as the difference in scalar value with the second highest maximum adjacent to it. The multi-saddle is canceled by reversing gradient paths that originate from it towards all maxima, except for the highest maximum, also called the surviving maximum.

\myparagraph{Saturated persistence directed simplification.}
While persistence directed cancellation removes saddles whose persistence falls below a given threshold, it does not remove long and potentially unintuitive arcs that connect spatially distant maxima via a saddle. These additional arcs often do not represent structural features. In addition, they cause clutter and occlude potentially interesting features. We propose the use of saturated persistence directed extremum graph simplification~\cite{correa2011topological}, which prunes such arcs. 

We apply all three operations described above to obtain simplified extremum graphs for various scientific datasets as shown in Figure~\ref{fig:simplified-extremum-graph}.


\section{Hybrid GPU-CPU parallel computation}
\label{sec:hybridparallel}
%
The parallel algorithm described in the previous section works when the dataset together with the associated data structures fit within the GPU and main memory. The GPU memory is often much smaller and determines the size of the data that can be processed. We now describe an extension of both steps of the gradient path tracing algorithm to handle datasets that are too large to fit within the GPU memory. We do assume, however, that the data fits within the main memory. In order to classify  critical points in large datasets, we view the dataset as an array of scalar field values in row major order. Thus, each vertex of the grid is mapped to an index of this array. We partition the array into contiguous subarrays called \emph{blocks}. \highlightadd{These contiguous blocks partition the $n$-dimensional dataset along one of its dimensions, for example along the y-axis for 2D data and along the z-axis for 3D data. A block is typically adjacent to two blocks, except for the first and last block which are adjacent to exactly one block.} 
The domain is partitioned such that each block fits in the available GPU memory during runtime. The size of a block is computed based on a linear regression model that predicts the number of vertices that can be accommodated in the GPU. The linear regression model is constructed from a series of experiments on datasets of different sizes executed on different GPUs.

Critical point classification of vertices that lie on the boundary of a block requires access to their neighboring vertices. So, we append each block with a set of ghost vertices constituted by vertices from the link of the boundary. These ghost vertices are not classified into regular or critical while processing the block. Critical point classification of all vertices in a block is computed on the GPU and the results are transferred to main memory. This process is repeated for all blocks in sequence. If multiple GPUs are available, then multiple blocks could be processed concurrently. \parext currently does not utilize multiple GPUs.
\newcommand{\hybridimagescale}{0.155}
\begin{figure}[!ht]
    \centering
    \begin{tabular}{c}
        \subcaptionbox{Sequential, GPU followed by CPU execution.}{\includegraphics[scale=\hybridimagescale]{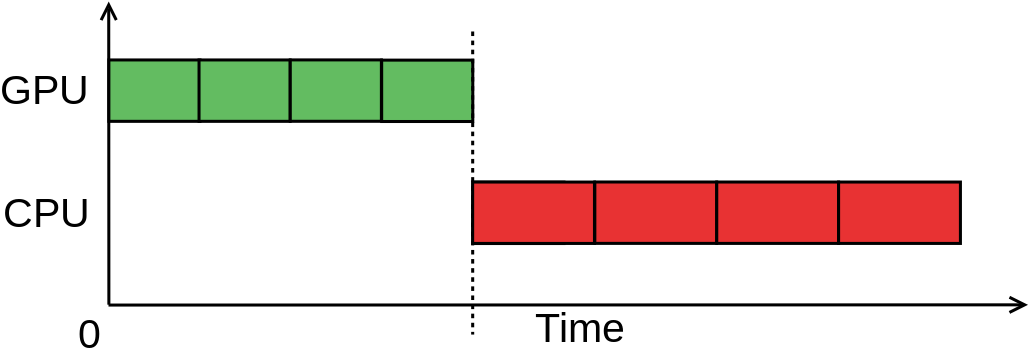}} \\
        \subcaptionbox{Hybrid GPU-CPU mode.}{\includegraphics[scale=\hybridimagescale]{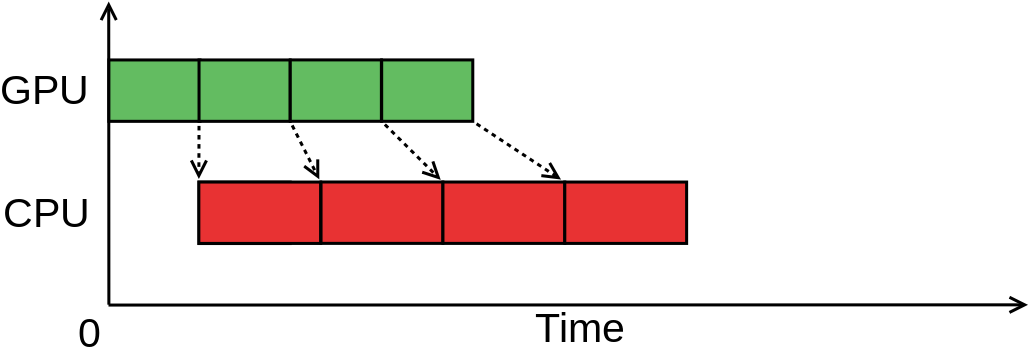}} \\
    \end{tabular}   
    \caption{Hybrid GPU-CPU parallel computation for a dataset partitioned into 4 blocks. In the sequential execution model~(a), point classification step (green) processes the blocks one by one on the GPU. Gradient path tracing~(red) for the blocks executes on the CPU after point classification is complete for all blocks. The CPU cores lie idle, waiting for the completion of critical point classification for all blocks on the GPU. In the hybrid GPU-CPU mode~(b), gradient path tracing within the first block begins soon after it is processed by the GPU. The transfer of control from GPU to CPU for a block is indicated by arrows. The base of the arrow lies at the finishing time for a block on GPU and the tip lies at the time when that block is taken up for processing on the CPU.}
    \label{fig:hybrid_gpu_cpu_timing_diagram}
\end{figure}
 
We could employ a simple approach for parallel gradient path tracing after all vertices are processed in the first step. However, this leads to a waste of CPU resources while the GPU is executing the first step. An optimal approach utilizes the CPU simultaneously for gradient path tracing within blocks that are already processed in the GPU. However, this path tracing is non-trivial. The gradient path tracing is indeed executed within each block but the task needs to be temporarily paused when the path reaches the block boundary and attempts to exit the block.  Figure~\ref{fig:hybrid_gpu_cpu_timing_diagram} illustrates how such an approach reduces the idle CPU time. 

 Gradients paths that need to exit a block are recorded and their tracing is resumed once the adjacent block is processed by the GPU.
\highlightadd{Path tracing in this hybrid mode is performed by maintaining a list of partially traced paths, whose tracing could not be completed because it entered a block that is not yet processed by the GPU to classify the critical points. A partial path is stored as a triple -- the origin saddle, the first vertex on the path, and the last visited vertex. The first vertex is required to distinguish between all paths that originate from the saddle. For each partial path, the last visited vertex is updated when path tracing is paused at a block boundary. Path tracing is resumed when the adjacent block containing the vertex is processed by the GPU. The tracing of a specific path continues until a maximum is reached and the overall tracing process terminates when the list of partial paths is exhausted. Thus, each vertex on a saddle-maximum path is visited once, which implies that the worst case running time is determined by the total number of vertices across all traced paths. The space required for path tracing is equal to the maximum size of the partial paths list.}

The above method can be further optimized by noticing that at any instant the collection of blocks processed by the GPU thus far can be viewed as a single but larger contiguous block. This is true if the blocks are processed in sequence. We fuse the blocks, iteratively including the next block after point classification is complete. This fuse step reduces the number of boundary crossings that may occur during gradient path tracing by simply reducing the number of possible boundaries to one, thus allowing for a majority of threads to complete tracing. By nature of this scheme, the extremum graph is computed without the need of any further processing, once gradient path tracing is completed for the last block.

\section{The \parext library}
\parext (\href{https://bitbucket.org/vgl_iisc/tachyon}{\url{bitbucket.org/vgl_iisc/tachyon}}) is a C++ software library that enables efficient computation of the extremum graph by offloading major portions of the computation onto an Nvidia GPU that supports CUDA. We list below a few salient features of the software library:
\begin{itemize}
    \item Leverages GPU computation power via CUDA kernels to perform point classification, an embarrassingly parallel task.
    \item Written in C++14. Utilizes standard threading libraries and synchronization utilities to implement parallelism for most tasks that cannot be performed on GPUs, thereby ensuring ease-of-installation and portability.
    \item Generalized to work for $n$-dimensional scalar fields. Computes both maximum and minimum graph, as required.
    \item Automatically switches to a hybrid GPU-CPU mode, on the fly, for datasets that do not fit in GPU memory.
    \item Provides graph simplification and cancellation algorithms for downstream applications of the extremum graph.
    \item Utilizes code and memory optimization tricks, such as bit manipulation operations and caching, for effective use of memory.
    \item Provides a user friendly command line utility to specify input data format, choice of graph simplification, cancellation algorithms, etc.
\end{itemize}

The library is designed in a modular fashion. For example, it is possible to configure \parext to perform point classification on the CPU if NVidia GPUs are not available. On a multi-GPU setup, it allows a user to select which GPU to utilize for computations. Further, it can be extended to support multi-GPU computation. The library has only a small set of dependencies, making it a versatile tool that can be built and used on a wide variety of workstations.

\parext is bundled with a user friendly interface that enables a non-expert to use the library for computing the extremum graph. The user interface supports various desired functionalities such as computation of maximum and/or minimum graphs based on user requirement, graph cancellation and simplification operations, support for various data types for storing the scalar field ($8/16/32/64$-bit signed/unsigned integers and single/double precision floating point values).

\section{Experimental Results}
\label{sec:results}
Experiments were conducted on datasets of sizes varying from $64 \times 64 \times 64$ to $2048 \times 2048 \times 2612$. These datasets were obtained from the Open Scientific Visualization Dataset Repository~\cite{openscivis}. All experiments, unless stated otherwise, were run on a workstation with an Intel(R) Xeon(R) Gold 6258R CPU @ 2.70GHz 28 Cores/56 Threads,  640GB RAM, and an Nvidia GeForce RTX 3080 GPU with 10GB GDDR6X RAM. Runtimes were computed as an average over five samples after dropping the best and worst timings out of seven runs. GPU temperatures were also kept within optimal operating ranges of $< 85^{o}$~C. We have performed experiments to highlight different properties of \parext and to evaluate its performance. Table~\ref{tab:datasets_info} lists the various datasets used in the following experiments.

\begin{table}[h]
    \centering
    \begin{tabular}{r|r|r}
    \textbf{Dataset} & \textbf{Size} & \textbf{\#Critical Pts} \\ \hline
    Nucleon & 41$\times$41$\times$41 & 421 \\
    Silicium&98$\times$34$\times$34&548 \\
    Neghip&64$\times$64$\times$64&1646 \\
    Fuel&64$\times$64$\times$64&296 \\
    Hydrogen&128$\times$128$\times$128&11370 \\
    Shockwave&64$\times$64$\times$512&991 \\
    Lobster&301$\times$324$\times$56&347207 \\
    Head Mri Ventricles&256$\times$256$\times$124&1719145 \\
    Engine&256$\times$256$\times$128&501799 \\
    Statue Leg&341$\times$341$\times$93&454011 \\
    Boston Teapot&256$\times$256$\times$178&101967 \\
    Skull&256$\times$256$\times$256&1877366 \\
    Foot&256$\times$256$\times$256&788482 \\
    Aneurism&256$\times$256$\times$256&60767 \\
    Bonsai&256$\times$256$\times$256&210570 \\
    Mrt Angio&416$\times$512$\times$112&5146370 \\
    Heptane Gas& 302$\times$302$\times$302 & 61212 \\
    Stent&512$\times$512$\times$174&3083028 \\
    Pancreas&240$\times$512$\times$512&7215261 \\
    Backpack&512$\times$512$\times$373&7066081 \\
    Magnetic Reconnection&512$\times$512$\times$512&31147751 \\
    Zeiss&680$\times$680$\times$680&3209058 \\
    Marmoset Neurons&1024$\times$1024$\times$314&56384791 \\
    Stag Beetle&832$\times$832$\times$494&852383 \\
    Pawpawsaurus&958$\times$646$\times$1088&83014508 \\
    Spathorhynchus&1024$\times$1024$\times$750&47569170 \\
    Kingsnake&1024$\times$1024$\times$795&32483997 \\
    Chameleon&1024$\times$1024$\times$1080&49325513 \\
    Beechnut&1024$\times$1024$\times$1546&159472969 \\
    Richtmyer Meshkov&2048$\times$2048$\times$1920&31585707 \\
    Woodbranch&2048$\times$2048$\times$2048&1520839571 \\
    3D Neurons & 2048$\times$2048$\times$2384&1913765935 \\
    Pig Heart&2048$\times$2048$\times$2612&906730898 \\
    \hline
    \multirow{2}{*}{Schwefel 3D} & 128$\times$128$\times$128 to & \multirow{2}{*}{1688} \\
    & 1024$\times$1024$\times$1024 &  \\
    \end{tabular}

    \caption{Datasets used in the computational experiments~\cite{openscivis}, their size, and the number of critical points. The Schwefel dataset~\cite{schwefel1981numerical} is resampled on domains of different sizes.}
    \label{tab:datasets_info}
\end{table}

\subsection{Internal memory computation}
\begin{figure}[ht]
    \centering
    \includegraphics[scale=0.2425]{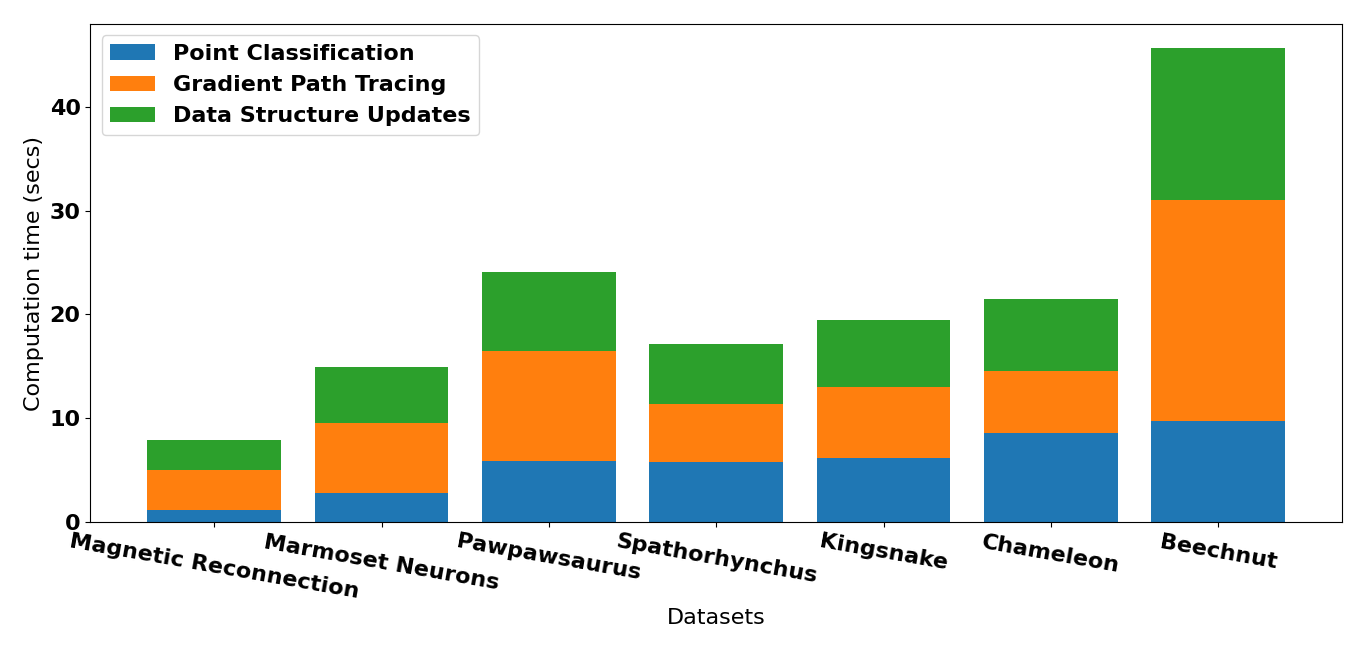}
    \caption{Extremum graph computation for datasets that fit in GPU memory. The time taken increases with size of dataset and naturally depends on the number of critical points.}
    \label{fig:res_internal}
\end{figure}

We first evaluate the runtime performance of the algorithm for datasets that fit within the GPU memory. In this experiment, we compute the wall clock time for computing the extremum graphs on a large collection of datasets, see Table~\ref{tab:datasets_info}.
These computations did not require the hybrid GPU-CPU execution. Figure~\ref{fig:res_internal} shows the total runtime and the split up between the two major steps and the time for data structure updates following the computation. Datasets that are smaller in size compared to Magnetic Reconnection require under $2.3$s. 
The figure shows runtimes only for larger datasets that fit in memory. Our key observations are:
\begin{itemize}
    \item Time required for critical point classification on the GPU scales linearly with the domain size.
    \item Time required for gradient path tracing on the CPU scales roughly linear with the number of identified saddles.
\end{itemize}

We also perform this experiment for higher dimensional data \highlightadd{by sampling the Schwefel function~\cite{schwefel1981numerical} on 3-to-6 dimensional domains of varying sizes}. Results are shown in Table~\ref{tab:higher_dimension_schwefel}. We again observe an increase with size of data, particularly the time taken for critical point classification.
\begin{table}[h]
    \centering
    \begin{tabular}{r|r|r|r}
    \textbf{Dataset} & \textbf{Size} & \textbf{\#Critical Pts} & \multirow{2}{2cm}{\textbf{Computation time (in secs)}} \\
    & & & \\
    \hline

    \multirow{3}{*}{Schwefel 4D} & $32^4$ & 8449 & 0.20 \\
    & $64^4$ & 15072  & 0.71 \\
    & $128^4$ & 15072 & 7.84 \\

    \hline

    \multirow{3}{*}{Schwefel 5D} & $16^5$ & 9525 & 0.33\\
     & $32^5$ & 62167 & 3.87\\
     & $64^5$ & 128808 & 110.01\\

    \hline

    \multirow{3}{*}{Schwefel 6D} & $16^6$ & 46345 & 9.13 \\
     & $24^6$ & 444241 & 104.46\\
     & $32^6$ & 444241  & 432.54\\

    \end{tabular}

    \caption{Extremum graph computation for higher dimensional scalar fields. A significant fraction of time was taken for the critical point classification step due to the increased neighborhood size.}
    \label{tab:higher_dimension_schwefel}
\end{table}

\subsection{Hybrid GPU-CPU hybrid computation}
\begin{figure}[h]
    \centering
    \includegraphics[scale=0.2425]{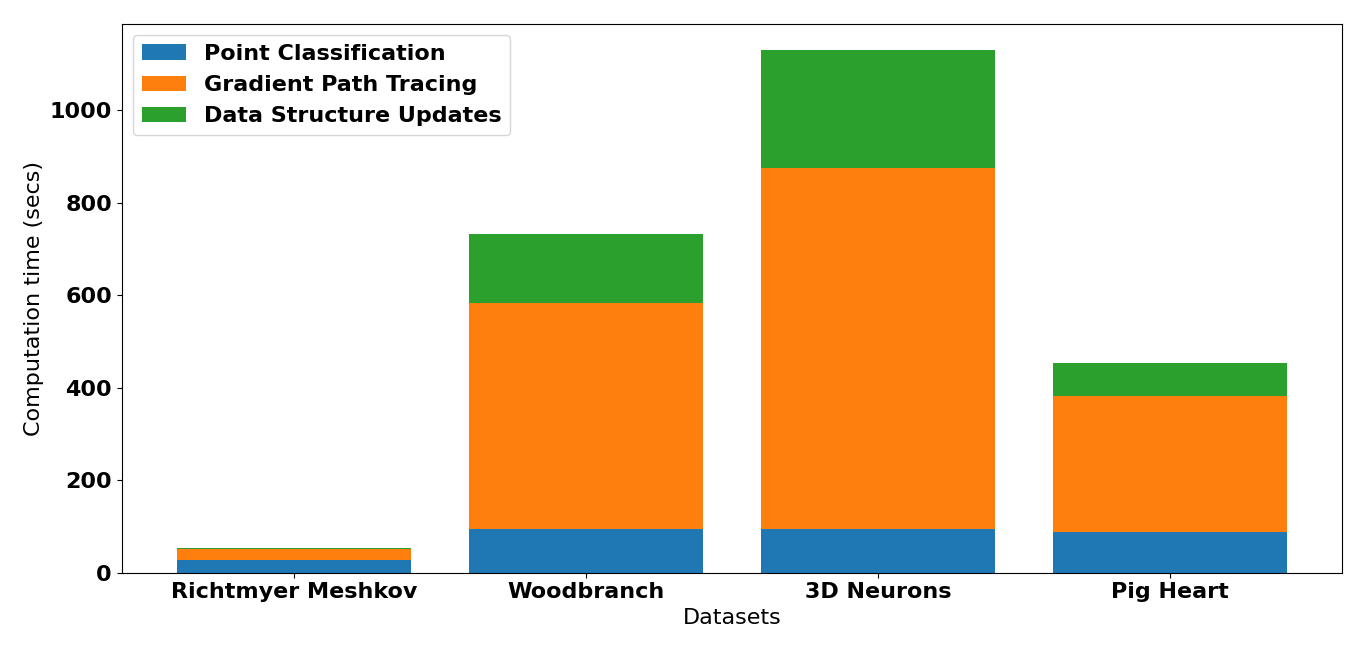}
    \caption{Performance for large datasets that do not fit in GPU memory. Runtime broadly depends on data size. Pig Heart has a small number of critical points, which explains the smaller runtime. Gradient path tracing constitutes a significant fraction of total runtime in all cases.}
    \label{fig:res_external}
\end{figure}

In this experiment, we choose the large datasets that do not fit in GPU memory and required partitioning into multiple blocks. Figure~\ref{fig:res_external} shows the runtime results for these datasets. We make the following observations:
\begin{itemize}
    \item Critical point classification time is nearly identical for all datasets because they have similar size.
    \item The gradient path tracing step consumes a large fraction of total runtime.
    \item An additional factor that influences gradient path tracing time is the number of paths that cross block boundaries. Such paths carry a computational overhead.
\end{itemize}

\subsection{GPU scaling - varying number of cores}
\begin{figure}[h]
    \centering
    \includegraphics[scale=0.125]{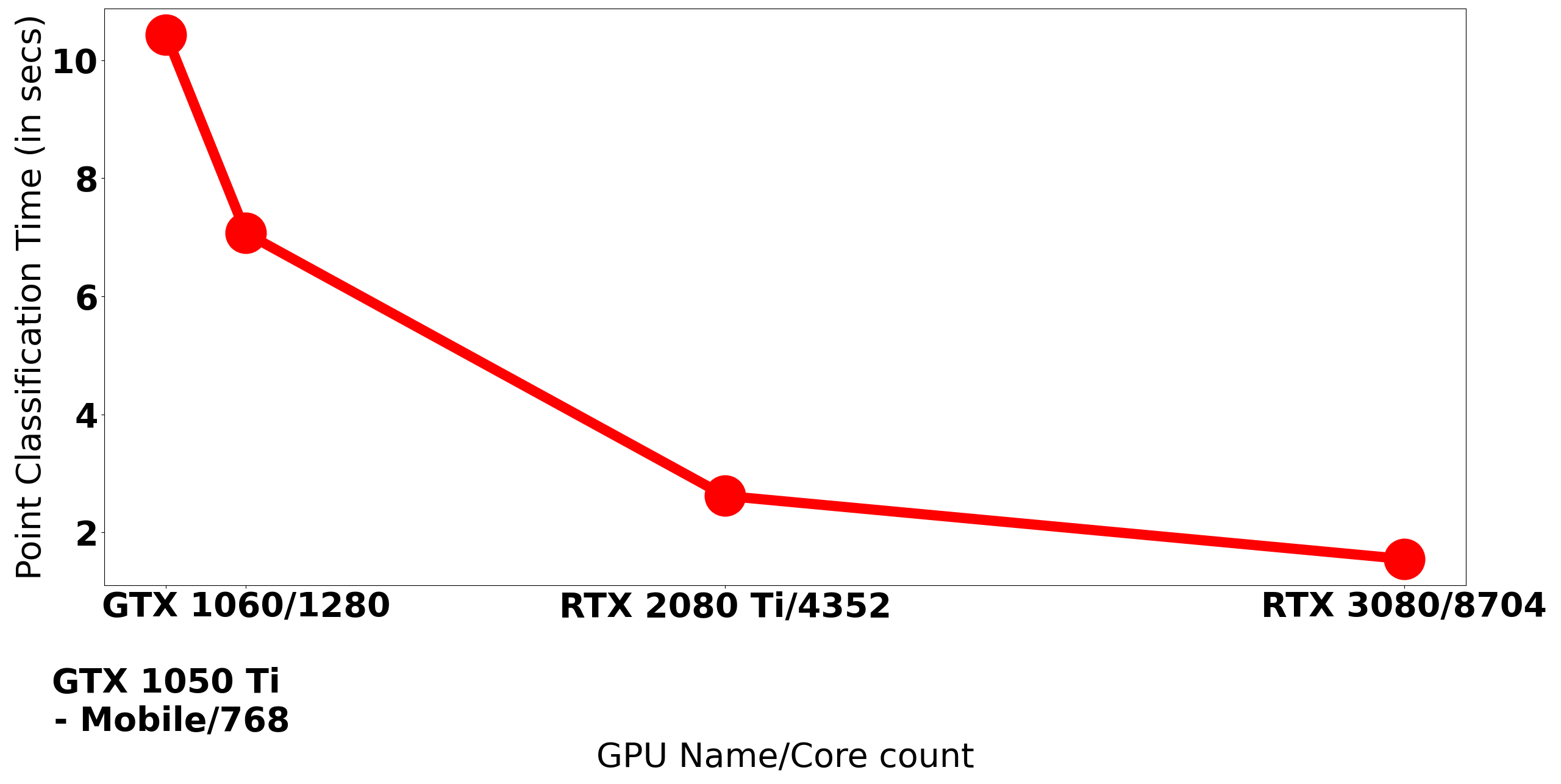}
    \caption{GPU scaling results.}
    \label{fig:res_gpu_scaling1}
\end{figure}

We evaluate the performance of the critical point classification step across different GPUs. These experiments are performed on the dataset Zeiss. The aim is to understand whether the computation scales to the number of available compute units on the GPU. Total computation time is not compared in this experiment because it is strongly influenced by the number of CPU cores and additional hardware features. The results are shown in Figure~\ref{fig:res_gpu_scaling1}. We observe a monotonically decreasing trend with a good slope, indicating efficient usage of the GPU cores. 

\subsection{GPU scaling - varying domain size}
\begin{figure}[h]
    \centering
    \includegraphics[scale=0.125]{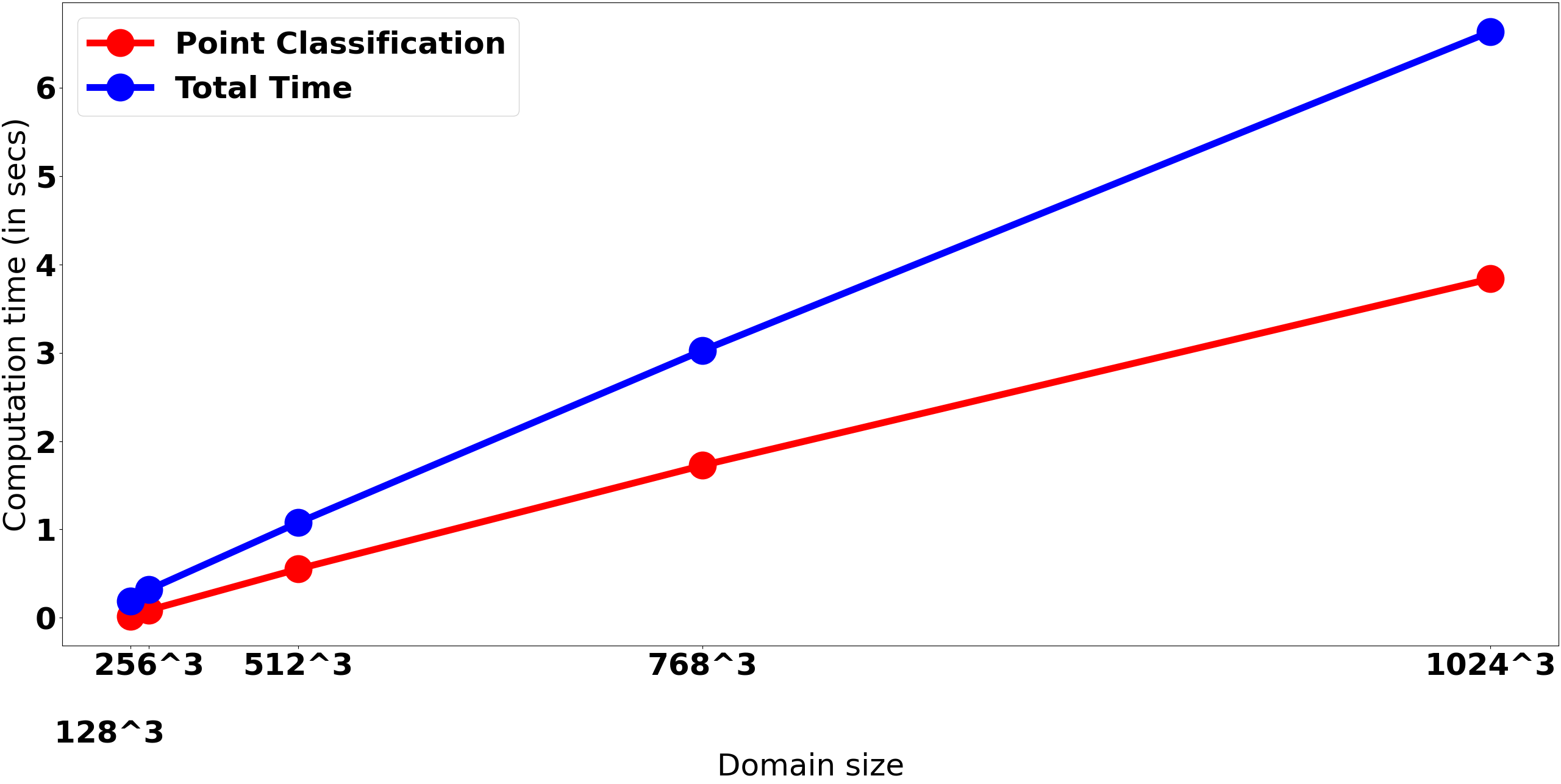}
    \caption{GPU scaling study for varying domain size. Experiments on datasets obtained by sampling the 3D Schwefel function~\cite{schwefel1981numerical} over different domain sizes reveal a linear relationship between computation time and domain size.}
    \label{fig:res_gpu_scaling2}
\end{figure}

This experiment aims to establish how well \parext utilizes the available GPU compute resources. We study how performance varies upon increasing the domain size while fixing the number of critical points. The Schwefel function~\cite{schwefel1981numerical} is sampled on a 3D domain. The data, originally available over a $500 \times 500 \times 500$, is sampled to obtain datasets over domains whose sizes range from $128 \times 128 \times 128$ to $1024 \times 1024 \times 1024$. Across all these sampled functions, the number of critical points and gradient paths do not change thereby allowing a study of the variation of GPU computation time with domain size. All GPU cores are used to compute the extremum graph for this dataset available at different resolutions. The results are shown in Figure~\ref{fig:res_gpu_scaling2}. We observe a strong linear dependency on domain size.

\subsection{CPU scaling - varying CPU threads}
\begin{figure}[h]
    \centering
    \includegraphics[scale=0.24]{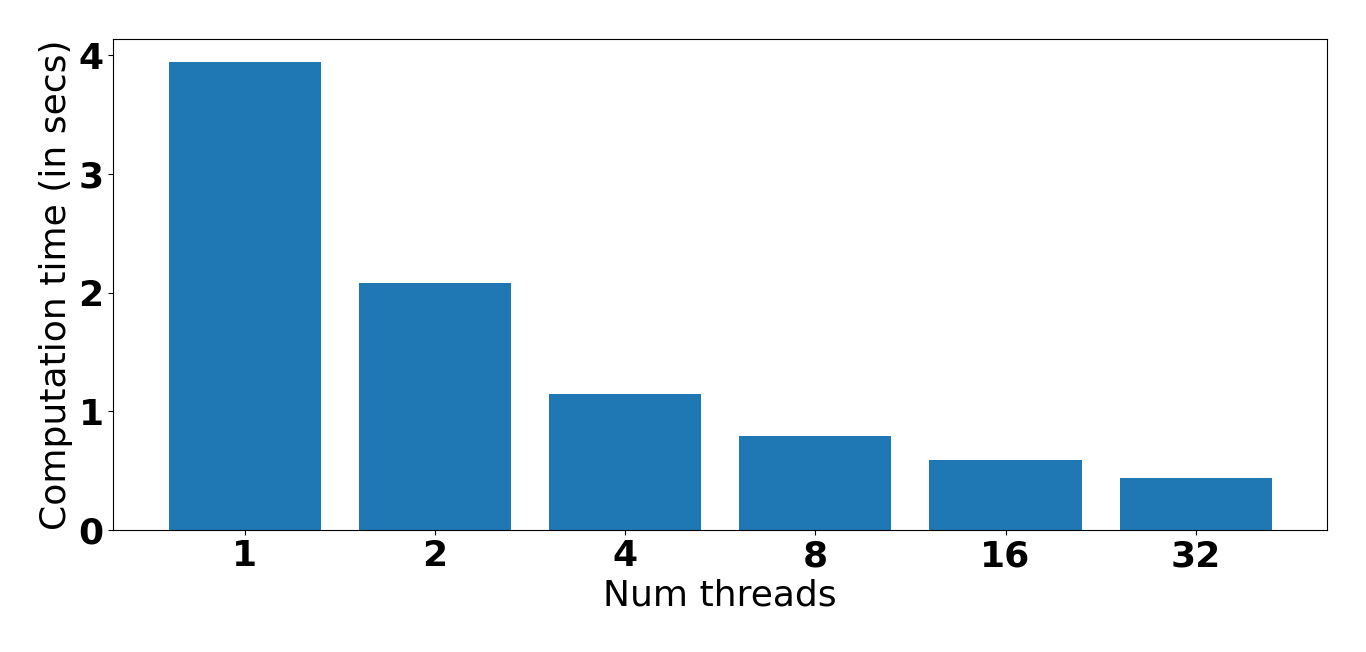}
    \caption{CPU scaling study for varying number of CPU threads. Experiments on the dataset Zeiss show good scaling up to 8 threads.}
    \label{fig:res_cpu_scaling2}
\end{figure}

Next, we study scaling behavior by varying the number of CPU threads. The aim is to study how well \parext  utilizes the available parallelism provided by the CPU. In the ideal case, we expect the computation time to reduce by a factor of two on doubling the number of threads. The results of the experiment are shown in Figure~\ref{fig:res_cpu_scaling2}. We observe that the performance improvements continue up to 8 threads and taper off afterwards.

\subsection{Performance comparisons}
We now describe results of runtime comparisons against TTK and pyms3d. Both TTK~\cite{ttksoftware} and pyms3d~\cite{mscsoftware2017} support computation of the complete Morse-Smale complex. 
\highlightadd{The key difference from \parext is that they employ a discrete Morse theory based approach. As a first step, they compute a discrete gradient field on the GPU to locate all critical cells. In contrast, \parext uses a lower link based critical point classification to locate nodes of the extremum graph. They employ a parallel root finding algorithm to compute the entire descending 3-manifold of all extrema and hence compute saddle-maxima arcs. In contrast, \parext employs gradient path tracing from each saddle to compute the arcs.
The gradient field and saddle-extrema 
 arc computation in \gmsc~\cite{subhash2022gpu} and pyms3d are identical. The only difference between the two implementations is that \gmsc uses CUDA whereas pyms3d uses OpenCL. }

\begin{figure}[h]
    \centering
    \includegraphics[scale=0.12]{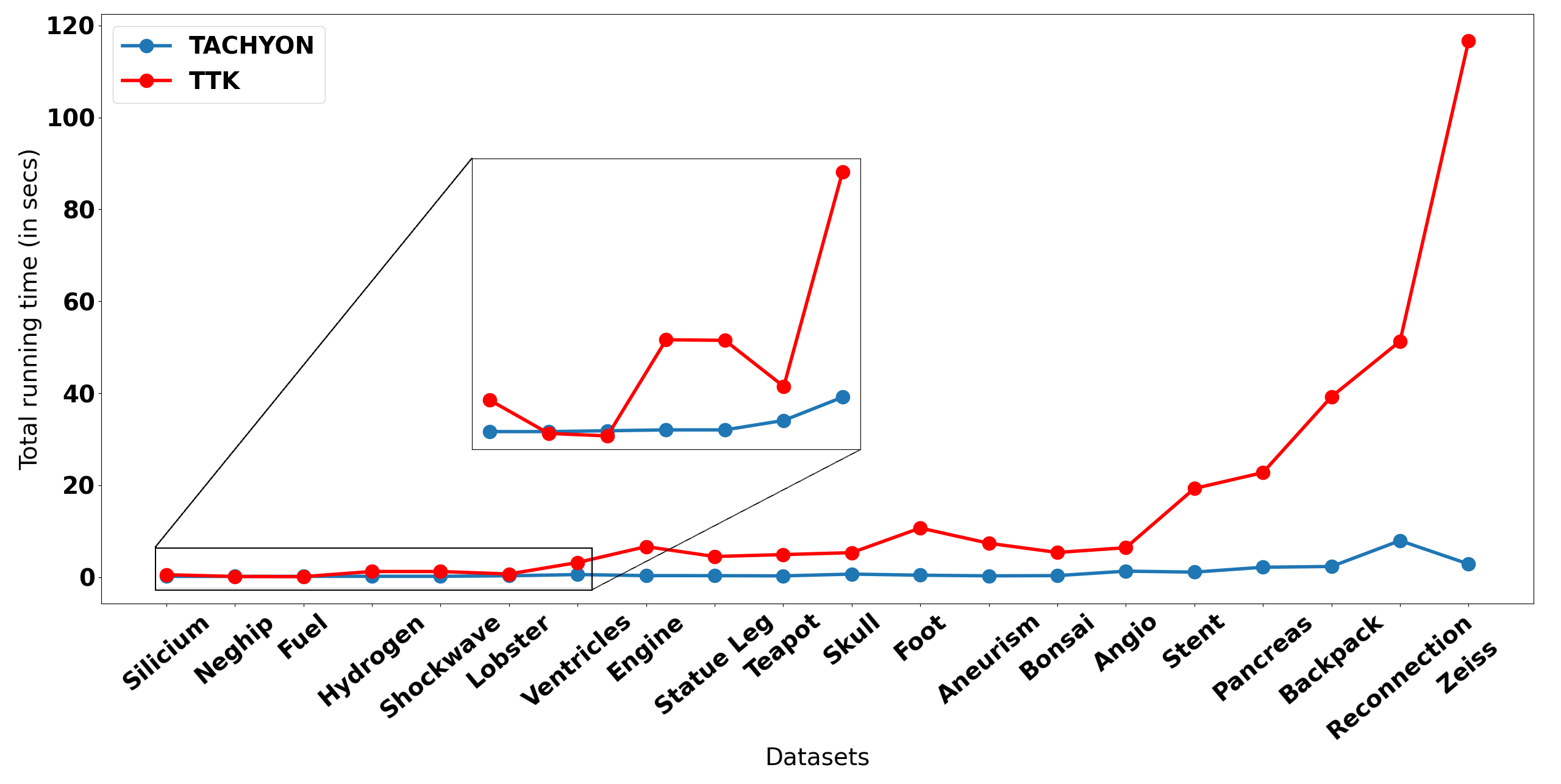}
    \caption{Runtime comparison between \parext and TTK\cite{ttksoftware}. \parext performs consistently better, often several orders of magnitude faster.}
    \label{fig:ttk_comparison}
\end{figure}

\begin{figure}[h]
    \centering
    \includegraphics[scale=0.2425]{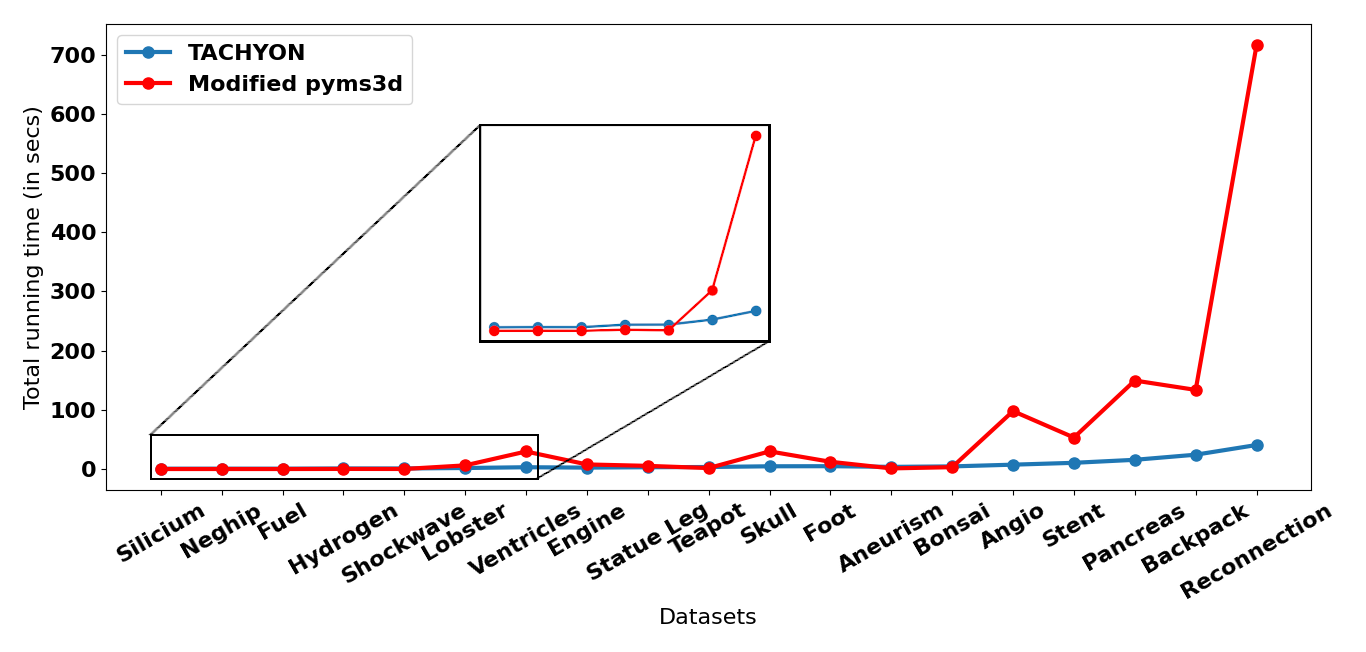}
    \caption{Runtime comparison between \parext and pyms3d~\cite{mscsoftware2017}. \parext performs consistently better, often several orders of magnitude faster.}
    \label{fig:pyms3d_comparison}
\end{figure}

In order to ensure a fair comparison, we configure pyms3d and TTK  to selectively compute the ascending 1-manifolds of 2-saddles and hence output the extremum graph. The results show that our implementation is able to compute extremum graphs faster. In some instances, both TTK and pyms3d fail due to system (GPU/main memory) resource limitations. We find that \parext performs better, both in terms of running time and utilization of system resources. Figures~\ref{fig:ttk_comparison} and~\ref{fig:pyms3d_comparison} plot the runtimes for increasing dataset sizes. The running time for \parext rises slower compared to TTK and pyms3d. Further, pyms3d ran out of memory and could not complete execution for larger data sizes. The comparison with pyms3d was performed on a workstation with an Intel(R) Core(TM) i5-4590 CPU @ 3.30GHz 2 Cores/4 Threads, 16GB RAM, and an Nvidia GeForce GTX 1060 GPU with 6GB GDDR5 RAM.

\subsection{Simplification}
\begin{figure}[h]
    \centering
    \begin{tabular}{c}
         \includegraphics[width=0.5\linewidth]{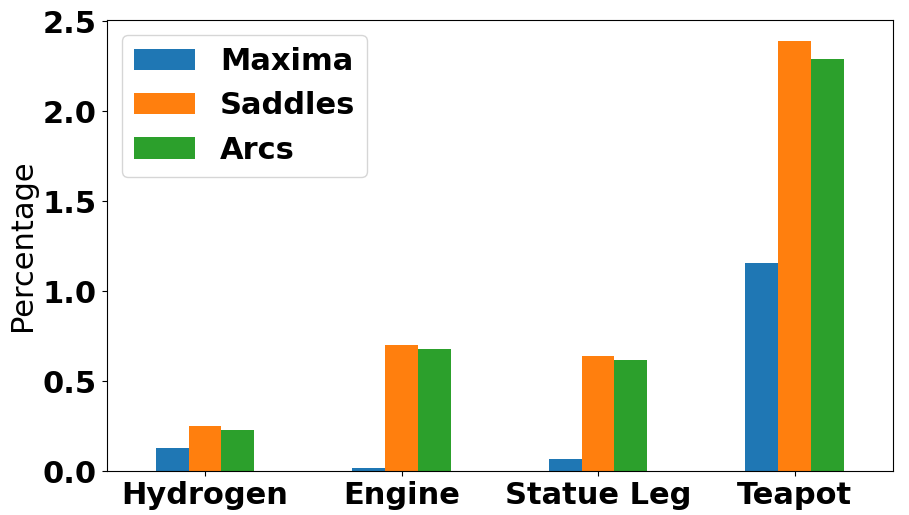}
         \includegraphics[width=0.5\linewidth]{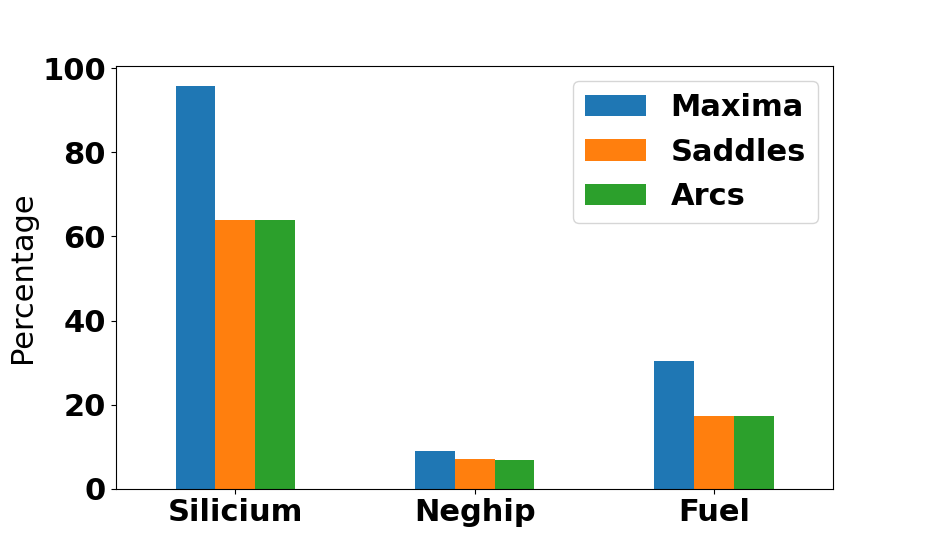}
    \end{tabular}
    \caption{Statistics on the fraction of extremum graph nodes and arcs that survive post simplification. The number of surviving extrema, saddles, and arcs after applying the three simplification operations in sequence are shown as a fraction of the corresponding numbers in the unsimplified extremum graph.}
    \label{fig:cancellation_results}
\end{figure}

Finally, we study the ability of the simplification operations to remove noise and clutter. We perform three operations on various datasets with a fixed parameter set: edge bundling, persistence directed cancellation with $5$\% threshold and saturated persistence directed simplification with high and low thresholds $p_{lo}=5\%$ and $p_{hi}=95\%$. Figure~\ref{fig:simplified-extremum-graph} shows the simplified extremum graphs. We also study statistics of the size of extremum graph before and after simplification, see Figure~\ref{fig:cancellation_results}. In some noisy datasets, a small threshold already removes a large fraction of nodes and arcs whereas in other datasets that contain a smaller number of features, the reduction is not significant. 

Across all results shown in Figure~\ref{fig:cancellation_results}, we observe that the number of arcs is a little over twice the number of saddles, although these absolute numbers are not depicted in the graph plot. This is expected, because all saddles have at least two outgoing arcs. 
%

\section{Conclusions and Future Work}
\label{sec:conclusions}
We have developed \parext, a dimension independent, scalable extremum graph computation library that also supports a hybrid GPU-CPU mode to process large datasets. This software library will be released in the public domain for use by the community. We also describe three methods for simplifying the extremum graph with the aim of identifying and removing noise, resulting in a clutter free graph that is more amenable for further analysis. Computational experiments demonstrate good running times and scaling properties.

Future work includes incorporating support for generic non-uniform grids and an extension to a distributed setting where the dataset does not fit in system memory. For the latter problem, an approach similar to the hybrid GPU-CPU mode where different processors work on different blocks of the same dataset may work but the challenge is to reduce the communication necessary to consolidate the output.

\section*{Acknowledgments}
This work is partially supported by a Swarnajayanti Fellowship from the Department of Science and Technology, India (DST/SJF/ETA-02/2015-16), an SERB grant CRG/2021/00527, MoE Govt. of India, and a Mindtree Chair research grant).

\bibliographystyle{eg-alpha-doi} 
\bibliography{references.bib}

\end{document}